\documentclass[]{interact}

\usepackage[caption=false]{subfig}

\usepackage[numbers,sort&compress]{natbib}
\bibpunct[, ]{[}{]}{,}{n}{,}{,}

\usepackage{graphicx}%
\usepackage{multirow}%
\usepackage{amsmath,amssymb,amsfonts}%
\usepackage{amsthm}%
\usepackage[title]{appendix}%
\usepackage{xcolor}%
\usepackage{textcomp}%
\usepackage{manyfoot}%
\usepackage{booktabs}%
\usepackage{algorithm}%
\usepackage{algorithmicx}%
\usepackage{algpseudocode}%
\usepackage{listings}%

\usepackage{tikz-cd}

\usepackage[numbers,sort&compress]{natbib}
\bibpunct[, ]{[}{]}{,}{n}{,}{,}

\theoremstyle{plain}
\newtheorem{theorem}{Theorem}[section]

\theoremstyle{definition}
\newtheorem{definition}[theorem]{Definition}

\theoremstyle{remark}

\begin{document}
	
	\articletype{Preprint}
	\title{Measurement of quantum states and nonlocal correlations}
	
\author{
	\name{Gregory D. Scholes\textsuperscript{a}\thanks{CONTACT Gregory D. Scholes. Email: gscholes@princeton.edu }}
	\affil{\textsuperscript{a} Department of Chemistry, Princeton University, Washington Rd, Princeton 08544, New Jersey,U.S.A.}
}

\maketitle
\date{\today}
	
\begin{abstract}
	One of the major remaining questions concerning foundations of the quantum mechanical theory is how superpositions are measured such that definite outcomes are obtained. Here a resolution to that issue is proposed. It is shown how to obtain state vectors associated with measurements on the separated subystems of an entangled state, revealing how a single wavefunction encodes a special kind of set of statistical measurement outcomes. The result explains why measurements on the subsystems give definite outcomes and why measurements on one subsystem are correlated with those on the other. It is therefore concluded that the theory of quantum mechanics, without nonlinearities or \emph{ad hoc} assertions, can explain both the mechanism of state vector collapse and the reason for the paradoxical nonlocal correlations between separated subsystems. The theory also explains why nonlocal correlations do not require any kind of interaction between the subsystems.   
\end{abstract}


\newpage

\section{Introduction}

The postulates of quantum mechanics include the foundational concept that an isolated physical system is described by a wavefunction $\Psi$, or state vector $|\Psi\rangle$ belonging to a Hilbert space $\mathcal{H}$. Experimental observations are tied to the meaurement postulates\cite{vonNeumann, Peres, OmnesBook}, which say the probability of obtaining an eigenvalue $a_n$ of a Hermitian operator  $\boldmath{A}$ is given by the amplitude squared $P(a_n) = |\langle a_n | \Psi \rangle|^2$, and that if a measurement gives the outcome $a_n$, then the state of the system immediately after the measurement is the eigenspace associated with $a_n$. This latter statement implies the `collapse' rule, which intuitively resolves the problem of obtaining a physical measurement outcome from a superposition state. Further, it guarantees that a subsequent measurement on the state also gives eigenvalue $a_n$. However, this apparently satisfactory result is difficult to explain based on the theory of quantum mechanics, and is central to the so-called quantum measurement problem\cite{quantumworld, Schlosshauer, Gomes2026, Ghirardi1988}. An accessible historical account can be found in Ref. \cite{Becker}.

Zwirn gives a nice summary of the main issue in a recent paper\cite{Zwirn2026} that we summarize here. Consider a quantum state $|\psi \rangle_s = \frac{1}{\sqrt{2}}(|0 \rangle_s + |1 \rangle_s)$. Now perform a measurement in the basis $\{ |0 \rangle_s, |1 \rangle_s \}$. According to the postulate described above, the measurement outcome will be either $|0 \rangle_s$ or $|1 \rangle_s$, each with probability $\frac{1}{2}$. After the measurement the system will be in the measured state, rather than remaining as the superposition $|\psi \rangle_s $. A contradiction is found, however, if we consider the system interacting with the measuring apparatus according to the usual formalism of linear quantum mechanics. In that case we find that the output state is again a superposition state, which does not correspond to the definite outcomes announced by the postulate. Penrose puts this more bluntly\cite{Penrose1998}: ``On the face of it, the state vector reduction that is considered to take place in a measurement is simply in blatant contradiction to the process of unitary evolution." 

There is one additional ingredient to the quantum measurement problem, which is the measurement basis dependence of the outcomes. That is, the outcomes themselves are not all present in the state vector as a predetermined `list' that can simply be read out. The principle of superposition means that we are free to choose any measurement basis, and that choice decides the possible measurement outcomes\cite{Zurek2003, Mermin1993}. This fact is quite well understood and in the theory described below it will be seen to arise naturally. 

The big question with respect to  measurement of quantum (superposition) states is how are definite outcomes predicted by the linear theory of quantum mechanics? That is, what is the mechanism of state vector reduction? If that can be resolved within  the theory of linear quantum mechanics, then the contradiction outlined above is resolved. Researchers have sought a microscopic explanation for collapse of the state vector (state vector reduction) for a century\cite{OmnesBook, Laloe, Dickson, GriffithsBook, HughsBook, Schlosshauer, Zurek2003, Ghirardi1988, Deutsch1985}.  For example, Omnès in Chapter 2.6 of Ref. \cite{OmnesBook} discusses reduction of the wavepacket, also known as collapse of the wavefunction and concludes that ``... reduction appears as one of the major problems in the theory [of measurement]''. The central issue is that unitary evolution is incompatible with the notion of collapse, leading some researchers to propose that nonlinearities in the fundamental theory of quantum mechanics might be  needed to explain collapse as well as fundamental notions of decoherence\cite{GRW1988, GPR1990, Bassi2013}.  Other researchers have questioned whether collapse is even a feasible concept. Dickson, Chapter 2.2 of Ref. \cite{Dickson} writes ``What is it about \emph{measurement}, or \emph{observation} (as opposed to other physical processes), that causes such a collapse?'' Very recent work evidences that, even now, interpretation of measurements is an intriguing topic\cite{Zwirn2023, Grangier2002, Ney2025, Kumar2026, Vaidman2026}. 

In the present paper we summarize and expand recent advances that show how wavevector reduction (`collapse') can be explained  within the existing theory of linear quantum mechanics. There are three interconnected components to the resolution of the problem. First,  how do measuring devices measure superposition states? Second, what is the definition of measurements on separated subsystems comprising a composite state? Third, how do measurements give definite outcomes within the theory of quantum mechanics?

In addition, it is shown that not only can collapse of superpositions be explained within the existing theory of quantum mechanics, but that this same framework gives insight into the nature of nonlocal correlations. In a remarkable paper\cite{EPR} published over 90 years ago, Einstein, Podolsky and Rosen (EPR) highlighted the difference between quantum and classical correlations. One of the issues noted by EPR is that quantum mechanics makes the unsettling prediction that correlations found in the measurement of entangled states appear to involve inexplicable long-range interactions. Famously described as a `spooky action at a distance', this property is known as nonlocality and has to date defied physical explanation. Background details are found in many sources\cite{Peres, Laloe, Ballentine, CS1978, GHSZ1990, Mermin1985, ReidEPR, hidden10, Selleri}. For example, the possibility that local variables hidden in the theory explain nonlocal correlations\cite{hidden6} has been ruled out by the Bell inequality\cite{Bell1964} and related inequalities like the CHSH inequality\cite{Peres, CS1978}. These analyses predict that correlations in entangled states exceed the limits of those possible for any local variable theory\cite{Brunner}, which has since been verified by experiments\cite{Aspect1982, Aspect2015}. 

Yet, understanding the mechanism by which the correlations are revealed in joint measurements of the separated subsystems associated with entangled states remains unresolved. In particular, how do the correlations arise when the collapse model tells us that the superpositions on each subsystem reduce randomly to give a measured outcome? Or put another way, how does a measurement on subsystem A influence the reduction of the wavepacket on subsystem B, given that the two subsystems do not interact\cite{Werner2014}? Fundamentally, the issue is that `collapse' of the state vector---though an appealing and intuitive notion---to date has no mechanistic explanation in the linear theory of quantum mechanics. Although the presence of nonlocal correlations is beyond doubt, elucidating of the mechanism enabling these correlations, which are evident even over very large distances, would resolve a significant gap in the fundamental theory of quantum mechanics. Moreover, clarifying the matter might suggest new ways of exploiting or implementing nonlocal correlations. 

The paper is organized as follows. We start (Sec. 2) by defining a working theory for measurements that sets up this paper's perspective on a well-known issue---that the measurement apparatus gives an entangled output state. The foundation for our proposed solution is to understand what are separated subsystems and their implications for measurements (Sec. 3). In Sec. 4 we discuss the origin of definite measurement outcomes by summarizing the contextual phase model\cite{collapse} and defining the tensor product basis for composite quantum states. We summarize some key results from the prior paper\cite{collapse}.  We discuss how a single entangled wavefunction naturally gives statistical measurement outcomes, rather than the probabilistic outcomes suggested by collapse theory.  In Sec. 6 we describe how definite measurement outcomes are predicted and show that the framework follows the principles of the theory of contexts, systems and measurements\cite{CSM}. It is shown why superpositions appear to `collapse' to a definite measurement outcome.  In Sec. 7 we briefly discuss the connection between this contextual phase model, collapse, and nonlocality. In Sec. 8 we show the close relationship between measurement bases and contextual phases. It is then shown that contextual phase class does not influence the joint measurement correlations (i.e. contextual phase theory is not a hidden local variable model). The implications are discussed in Sec. 9. 

Following the convention in mathematics, we will often specify a state in a vector space without using Dirac notation. That is, we consider the notation $\psi$ equivalent to $| \psi \rangle$.

\section{Measurement of superpositions}

As a starting point, we need a clear definition for how measurement works with respect to the kinds of measurement problems we specialize to in this work. We thus define an ideal measuring device to have the following properties:

\begin{enumerate}
	\item The measuring device can be macroscopic and does not itself have to entangle with the state, although it may. Importantly, it facilitates appropriately projecting the input state to a superposition of the output states. A well-designed measurement apparatus maps basis states to physical outputs. The concept is a direct generalization of the quantum-mechanical lossless beam splitter\cite{KnightQOptics, beamsplitter}.  
	\item The output ports are equated to separated subsystems of the entangled output state. Measurements of the subsystems therefore give definite outcomes, according to the contextual phase theory\cite{collapse, sepsubs}. Microscopically, this property is controlled by the requirement that the quanta of output readings equals the quanta input into the device.
	\item A sequence of measurements gives a statistical representation of the input superposition, as ensured by the contextual phase theory\cite{collapse, sepsubs}. This distribution of measurement outcomes gives the standard probabilistic interpretation of a state.
\end{enumerate}

The concept for the measurement model is that an input state, for example $|\psi_{\text{in}} \rangle = \frac{1}{\sqrt{2}}(|0\rangle_s + |1\rangle_s)$, interacts with the measuring device according to the Schrödinger equation to give a system-device entangled state, for example $|\psi_{sm} \rangle = \frac{1}{\sqrt{2}}(|0\rangle_s |0\rangle_{m,A} + |1\rangle_s |1\rangle_{m,B})$. Here the subscript $m$ means measurement device and A, B indicate the explicit output port. The device does not have to be explicitly entangled, but it mediates the entanglement. The beam splitter example below explains this better. Contextual phase theory elucidates how we subsequently obtain a definite measurement outcome, evident as a meter reading, photon, etc. in either port A \emph{or} port B. Hence, this proposed approach to measurement resolves the apparent contradiction between the requirement for definite measurement outcomes and the quantum mechanical prediction for how the system interacts with a measuring device.

\begin{figure}[h]
	\centering
	\includegraphics[width=0.8\textwidth]{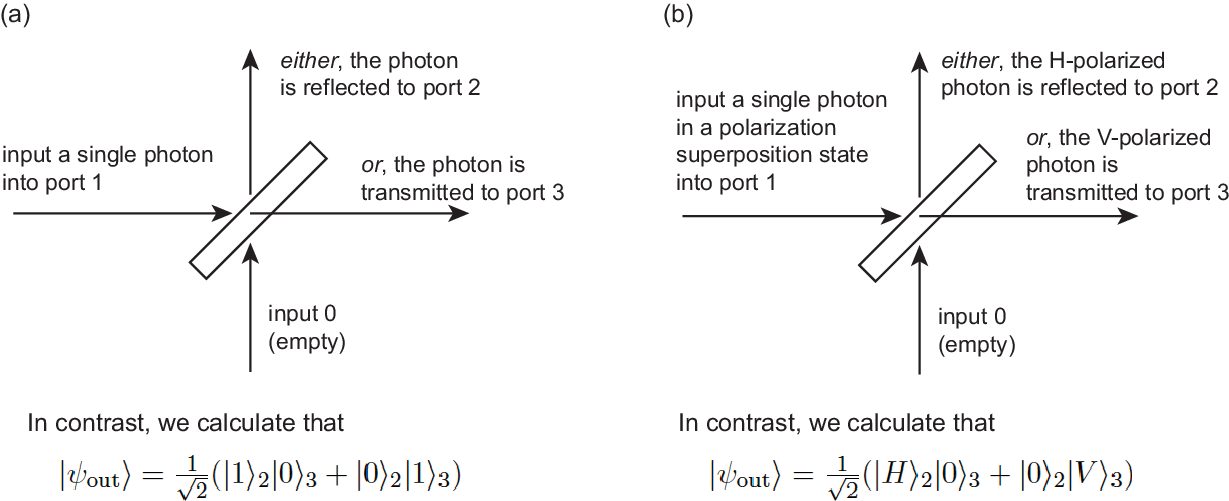}
	\caption{The beam splitter analogy for a general quantum measuring device that works by resolving a read-out into two separated subystems, ports 2 and 3. The system shows how the definite outcomes measured contrast to the output superposition states that are predicted by quantum mechanics. (a) Example of a single photon input into port 1. A single photon emerges, either from port 2 or port 3 with each definite outcome being equally likely. (b) A similar set-up, but with the single input photon being prepared in a polarization superposition. The beam splitter projects the superposition into either one of the two output ports, accordingly measuring horizontal polarization (H) or vertical polarization (V). Again, quantum mechanics predicts the output to be an entangled state.  }\label{fig1}
\end{figure}

We can relate this apparatus to the lossless beam splitter\cite{KnightQOptics, beamsplitter}, Fig. 1. A classical beam splitter works intuitively. An input beam with complex field amplitude $E_1$ enters the beam splitter from the left (call this port 1). The reflected part of the input beam gives field amplitude $E_2$, that emerges from port 2, while the transmitted fraction, $E_3$, emerges to the right, from port 3. We require that the intensity of the incident beam is distributed among the output beams, so that $|E_1|^2 = |E_2|^2 + |E_3|^2$. To translate this model to the quantum case we need to replace each $E_i$ by a corresponding annihilation operator $a_i$. In addition, we need to map a two-state input to a two-state output. In the case of the normal beam splitter, the second state of the input is the otherwise unused port into the beam splitter (port 0), which carries only the vacuum mode. Using scalars $r$ and $t$ to indicate the fraction of the beam entering port 1 that is reflected and transmitted, and similarly $r'$ and $t'$ for the fictional beam entering port 0, together with the reciprocity relations, one obtains the following relationship between annihilation (or creation) operators of the input and output ports:
\begin{equation}
	\begin{pmatrix}
		a_2 \\
		a_3  
	\end{pmatrix} 
	=
	\begin{pmatrix}
		t' & r \\
		r' & t  
	\end{pmatrix} 
	\begin{pmatrix}
		a_0 \\
		a_1  
	\end{pmatrix} 
	= U
	\begin{pmatrix}
		a_0 \\
		a_1  
	\end{pmatrix} 
	,
\end{equation}	
where $U$ is a unitary operator. We know that the photon must exit the beam splitter from \emph{either} port 2 \emph{or} port 3. However, using our quantum-mechanical model we predict an output state that is a superposition of these possibilities: $|\psi_{\text{out}} \rangle = \frac{1}{\sqrt{2}}(|1\rangle_2 |0\rangle_3 + |0\rangle_2 |1\rangle_3 )$.

By expanding this formalism to allow for photon polarization modes, we can model the polarizing beam splitter. If the input photon has an arbitrary linear polarization---a superposition of the two input modes $H$ and $V$, then that superposition is projected to a single definite outcome, which is an output photon detected from either the $H$ or $V$ output ports. This gives a concrete example of an input superposition being mapped to the basis states, which are physically distinct read-outs. 

It is computed that an input superposition of polarization modes $|\psi_{\text{in}} \rangle = \frac{1}{\sqrt{2}}(|H\rangle_1 |0\rangle_0 + |V\rangle_1 |0\rangle_0 )$ yields a superposition of the two possible output scenarios, the photon exits port 2 and there is a vacuum state in port 3, and \emph{vice versa}: $|\psi_{\text{out}} \rangle = \frac{1}{\sqrt{2}}(|H\rangle_2 |0\rangle_3 + |0\rangle_2 |V\rangle_3 )$. This entangled state appears at odds with the stipulation that we obtain a statistical distribution of definite outcomes from a sequence of experiments. However, contextual phase theory explains how measurements on the \emph{separated} subsystems---here the fact that the superposition is presented by the two output ports---do in fact yield definite outcomes without needing an \emph{ad hoc} notion of wavepacket reduction. The example of a regular lossless beam splitter is articulated in Ref \cite{collapse}. The present paper explains how definite outcomes are obtained. 

\section{Separated subsystems}

The starting point for understanding how measurements lead to definite outcomes is to define separated subsystems. Einstein spent much time thinking about this concept---which is perhaps a deeper idea than is widely appreciated. While trying to explain more clearly what he had in mind in the EPR paper, Einstein wrote on June 19, 1935 to Schrödinger\cite{FineBook}:
\begin{quotation}
	Separation is the claim that whether a physical property holds for one of the particles does not depend on measurements (or other interactions) made on the other particle when the pair is widely separated in space.
\end{quotation}

The way we would like to think about measurements on separated subsystems is that they provide a kind of projection of the composite state mediated  by a subsystem. That projection cannot be accomplished for general states in the tensor product space $\mathcal{H}_A \otimes \mathcal{H}_B$, where the basis for each subsystem has been converted to a new tensor basis for the composite system. Instead, we need to explain measurements on the separated subsystems in terms of expectation values of the subsystem states in their respective Hilbert spaces $\mathcal{H}_A$ and $\mathcal{H}_B$;  the vectors in these spaces are precisely those associated with separated subsystems.  In Ref. \cite{sepsubs}, we conveyed this concept by the following idea:

Measurements of separated subsystems:
Let $\Psi$ be a bipartite maximally entangled state specified in the tensor product space $\mathcal{T}(\mathcal{H}_A \otimes \mathcal{H}_B)$, where $\mathcal{H}_A$ is the Hilbert space of states for subsystem A and $\mathcal{H}_B$ is the Hilbert space of states for subsystem B. Measurement outcomes on the subsystems A and B when separated from $\Psi$ are the observables associated with projections of representatives of the equivalence class $[\Psi] \in \mathcal{F}(\mathcal{H}_A \times \mathcal{H}_B)$.

The concept that we unravel here is the notion of how a separated subsystem enables measurements of a composite state to be projected into the Hilbert space of that subsystem. In particular, we will explain that this is not a bilinear map that is linearized by the tensor product space. In other words, applying an operator like $T_A \otimes \mathbb{I}_B$ to a tensor product state (the standard strategy) does not do the job required.

It is stated in the following definition that the required projection happens from the free vector space $\mathcal{F}(U \times V)$ from which the tensor product space is defined (by the action of a quotient). The relationships and equivalences between vectors in these spaces will be discussed in a later section. See Appendix A for background on free vector spaces. For now all we need to know is that for any state in a tensor product space $\mathcal{T}(U \otimes V)$, there are representatives in the free vector space $\mathcal{F}(U \times V)$ from which the tensor product space derives. Here $U$ and $V$ are linear subspaces, for example, the Hilbert spaces $\mathcal{H}_A$ and $\mathcal{H}_B$.

States of the separated subsystems:
For a composite system with states in $\mathcal{T}(U \otimes V)$, separation of the subsystems is equivalent to the projections
\begin{eqnarray}
	\pi_U : U \times V \rightarrow U \\
	\pi_V : U \times V \rightarrow V \nonumber
\end{eqnarray}
where a projection map $\pi_U$ sends any pair $(u,v) \in U \times V$ to the first coordinate $u \in U$, and analogously for $\pi_V$. The map satisfies $\pi_U(x + y) = \pi_U(x) + \pi_U(y)$ for any $x, y \in U \times V$.

In general, such a projection map cannot be achieved when we work with tensor product spaces. To explain why, we start by briefly summarizing the relationship between a bilinear space $\mathcal{F}(U \times V)$ and its linearized counterpart, $\mathcal{T}(U \otimes V)$. General references include Ref \cite{Roman, TensorSpaces, DefantFloret}. The following conveys the approach given in Chapter 1 of Ref \cite{DFS}. 

The tensor product construction is motivated by looking for a linear space $W$ such that linear maps from $W$ to another linear space $Z$, $L(W; Z)$, coincide with the space of all bilinear maps (functions) on $U \times V$ to $Z$, denoted $B(U, V; Z)$. Finding such a connection provides a way of linearizing bilinear functions. As is well known, that is achieved by constructing the tensor product space $U \otimes V$. When we do this, the characteristic  properties of the bilinear map 
\begin{eqnarray*}
	B(u_1 + u_2, v) = B(u_1, v) + B(u_2, v) \\
	B(u, v_1 + v_2) = B(u, v_1) + B(u, v_2) \\
	B(\lambda u, v) = B(u, \lambda v) = \lambda B(u,v),
\end{eqnarray*}
where $\lambda$ is a scalar, $u, u_1, u_2 \in U$ and  $v, v_1, v_2 \in V$, are embedded in the properties of the tensors:
\begin{eqnarray*}
	(u_1 + u_2) \otimes v = (u_1 \otimes v) + (u_2 \otimes v) \\
	u \otimes (v_1 + v_2) = (u \otimes v_1) + (u \otimes v_2) \\
	(\lambda u) \otimes v = u \otimes (\lambda v) = \lambda (u \otimes v).
\end{eqnarray*}
We focus on these properties in a later section of the paper. 

The result we need is that the bilinear space $U \times V$ is related to its linearized counterpart $U \otimes V$ by the universal mapping property\cite{Roman, TensorSpaces, DFS}:
\[
\begin{tikzcd}[column sep=large, row sep=large]
	U \times V \ar[rd, "f "'] \ar[r, "t"] & U \otimes V  \ar[d, description, "\tau"'] \\
	& Z
\end{tikzcd}
\]
Here the map $\tau$ is a linear map, whereas $f$ and $t$ are bilinear maps. This establishes a natural isomorphism between $B(U, V; Z)$ and $L(W; Z)$ by the formula $\tau (u \otimes v) = f(u,v)$. 

Now we get to the reason for describing this universal mapping property, which is to explain why there is no canonical map $U \otimes V \rightarrow U$ analogous to the projection map on $U \times V$ described above. The reason is that the universal mapping property fails because $\pi_u(U \times V)$ is not a bilinear map. 

In Sec. 6 we show how to use the projection maps in practice and we will see how wavepacket reduction (collapse),  that occurs when a measurement is made, is explained as an interference phenomenon enabling a definite outcome for the chosen measurement conditions. First, we will establish an important result concerning how the phase of the superposition in an entangled state is transferred to the vectors projected to its separated subsystems.

\section{The basis for definite measurement outcomes}

In this section we provide the essence of our model for why measurements yield definite outcomes. The model, which we call the contextual phase model\cite{collapse}, proposes that measurements on separated subsystems are dictated by phase information that is hidden to operators acting on the space of composite states $U \otimes V$, but seen in the states projected to the separated subsystems. This additional phase information (contextual phase) leads to interference effects in measurements on the separated subsystems that reduce superpositions to definite outcomes. 

\subsection{The contextual phase model}

We start with a brief summary of our prior work\cite{collapse}, where we proposed the contextual phase theory for quantum measurement of separated subsystems. A way to associate entangled states in $\mathcal{H}_A \otimes \mathcal{H}_B$ to states that represent the appropriate local superpositions in the separated, non-interacting subsystems was posited. Those states lie in $\mathcal{H}_A$ and $\mathcal{H}_B$. 

The previous paper highlights the following facts, which give intuition for the approach that is explained rigorously in the following sections. In terms of vectors in the Hilbert space of subsystem A, phase is encoded in the coefficients that define any normalized state vector $\psi_A = \alpha|0\rangle + \beta|1\rangle$ with respect to our chosen basis. It is convenient to visualize the pair of complex coefficients $\alpha, \beta$, assuming $\psi_A$ is normalized, as a point on the sphere in four-dimensional real space, $\mathbb{S}^3$. The method shows the importance of accounting for the possible ways that the phase of the superposition associates to these vectors. The phase in the case of $\psi_A$ does not matter, it is an arbitrary global phase. Where the phase does matter is when it is the phase of a superposition state---that is, the phase prefactor for the second term of the linear combination for a composite state. This phase means that when we construct the tensor product space, we can visualize a \emph{set} of points in $\mathbb{S}^3_A \times \mathbb{S}^3_B$ that map to a \emph{single} vector in $\mathcal{H}_A \otimes \mathcal{H}_B$. We say that these classes of state vectors have different contextual phase factors. This observation is a key result of Ref. \cite{collapse}.

We gave a detailed derivation in another paper for how to obtain vectors in $\mathcal{H}_A$ and $\mathcal{H}_B$ corresponding to the separated subsystems from an entangled state\cite{sepsubs}. We then consider measurement outcomes, which depend as usual on the measurement performed. We have shown for any measurement there is a way that the state vector of the separated subsystem will appear to collapse and thus to yield a specific measurement outcome, without making any \emph{ad hoc} assumptions. It is important to note, this result concerns the measurements taking place on the subsystems---there can be no `collapse' or simplification of the state vector simply because of separation of the subsystems because then entanglement would be lost and we would not observe correlations that can exceed those possible for a classical system. A central result of the prior paper is\cite{collapse} is the following hypothesis.

State collapse for local measurement of isolated subsystems:
Collapse of a superposition local to an isolated subsystem A associated with an entangled state in $\mathcal{H}_A \otimes \mathcal{H}_B$,  that is, a superposition of vectors where each is an element of $\mathcal{H}_A$, can be reinterpreted as interference of a superposition of state vectors in the Hilbert space $\mathcal{H}_A$ caused by measurement of the isolated subsystem. According to the contextual phases randomly encoded in the state, the interference yields a measurement outcome in the chosen measurement basis. That is, while the contextual phases are random, the collapse induced by measurement is correlated among the separated systems.

According to the contextual phase theory, collapse of a single state vector is not random, but the outcome of collapse \emph{appears} to be random because the contextual phase class is statistically distributed in the ensemble of measurements on wavefunctions  presented as an equivalence class in the free vector space. That set of wavefunctions maps to a single wavefunction in the quotient space that defines the tensor product, as shown in the next subsection. Further, the collapse of a state vector for one subsystem in a composite state is correlated with the collapse of the state vector of the other subsystem, without the need for any kind of interaction after the subsystems separate. This is based on the way the phase of the superposition is associated  (or we can informally say `lifted') to the points in the covering space, and hence to the vectors of the separated subsystems in their respective Hilbert spaces. We must therefore associate the phase of the superposition explicitly with one of the subsystem state vectors. In the case of the Bell states, the phase is simply $+$ or $-$. 

We use a notation here for the basis vectors where $|-0\rangle \equiv -(|0\rangle)$, $|-1\rangle \equiv -(|1\rangle)$, $|0'\rangle \equiv \frac{1}{\sqrt{2}}(|0\rangle + |1\rangle)$, $|1'\rangle \equiv \frac{1}{\sqrt{2}}(|0\rangle - |1\rangle)$, $|-0'\rangle \equiv \frac{1}{\sqrt{2}}(-|0\rangle - |1\rangle)$, and $|-1'\rangle \equiv \frac{1}{\sqrt{2}}(-|0\rangle + |1\rangle)$. The derivation and implications for these representations of the entangled states is given in the following sections.

\subsection{Why a single wavefunction gives statistical-like measurement outcomes}

Einstein struggled with the nature of the wavefunction and how it is probed by measurements. The fact that a single wavefunction can yield various measurement outcomes according to how it `collapses' when measured is the issue which opens the possibility of a statistical interpretation for the wavefunction. Einstein thus wrote\cite{Einstein1936}:

\begin{quotation}
	The $\psi$ function does not in any way describe a condition which could be that of a single system; it relates rather to many systems, to ``an ensemble of systems'' in the sense of statistical mechanics.
\end{quotation}

Einstein considered that this statistical view suggests an incompleteness in quantum theory. Bohr, on the other hand advocated that the wavefunction describes a single system. Here it is shown that a single wavefunction that describes a composite quantum system intrinsically encodes a \emph{set} of outcomes for measurements on the separated subsystems. In this sense, the ensemble of measurement outcomes is `hidden in plain sight' and the probabilistic view of collapse is not required to explain measurements. That is, a sequence of identical $\Psi$ in $\mathcal{T}(\mathcal{H}_A \otimes \mathcal{H}_B)$ corresponds to an ensemble  of $\Psi$ in $\mathcal{F}(\mathcal{H}_A \times \mathcal{H}_B)$, and hence to a distribution of  outcomes for measurements on the separated subsystems. As we will see, there is still a probabilistic aspect to how the statistical distribution of definite measurement outcomes arises, which comes down to whether $\Psi$ has class 1 or 2 contextual phase with respect to the measurement projection.

Now define $\mathcal{F}(U \times V)$ as the free vector space over the field $\mathbb{C}$ on the set $U \times V$ (i.e. Cartesian product). It has elements that are finite $\mathbb{C}$-linear combinations of basis symbols $(u, v)$ and satisfies the universal property for maps from $U \times V$ into $\mathbb{C}$-vector spaces (see Sec. 3). 

Our results suggest a special kind of statistical basis for the wavefunction. Let $U_A$ and $V_B$ be linear state spaces associated with subsystems A and B respectively. An entangled state $\Psi \in U_A \otimes V_B$ denotes an equivalence class $[\Psi] \in U_A \times V_B$.

The assertion  follows immediately from the definition of tensor product\cite{TensorSpaces, WarnerBook, Roman}. Let $\mathcal{F}(U \times V)$ be the free vector space over the field $\mathbb{C}$ with basis $U \times V$. Let $\mathcal{R}$ be the subspace of $\mathcal{F}(U \times V)$ generated by all vectors of the form
\begin{eqnarray}
	r(u, w) + s(v, w) - (ru + sv, w) \\
	r(u, v) + s(u, w) - (u, rv + sw) \nonumber 
\end{eqnarray}
with $s, r \in \mathbb{C}$. These vectors are what we must identify as the zero vector in order to ensure bilinearity of the map from $U \times V$, which is the key requirement for defining the tensor product. We now define the tensor product by the quotient
\begin{equation}
	\mathcal{T}(U \otimes V) := \mathcal{F}(U \times V)/\mathcal{R} .
\end{equation}
Thus, by definition, each point in $\mathcal{T}(U \otimes V)$ corresponds to an equivalence class of points in $\mathcal{F}(U \times V)$. 

A key result of our recent prior work is that the tensor product space provides a way to group the vectors from the free vector space\cite{sepsubs}. Realizing that fact shows a way to think about how structure is hidden in the wavefunction $\Psi$ of an entangled state. $\Psi$ hides the representatives in $\mathcal{F}$ because it is defined in a quotient space. Whereas, the representatives are exposed when measurements are made on the separated subsystems, as will be shown below. This way of thinking about $\Psi$ enables understanding of why the measurement outcomes are definite, and why measurement outcomes are correlated among the subsystems.

The motivation for this particular kind of statistical interpretation is the intuition for a quotient space that we `group' elements together according to an equivalence relation, then we treat each group (i.e. each equivalence class) as a single point in the quotient space. See Sec. 4 of ref \cite{collapse}. Therefore, the wavefunctions $\Psi \in \mathcal{H}_A \otimes \mathcal{H}_B$ are all considered identical with respect to the action of operators on the states of the composite system (e.g. operators of the form $T_A \otimes T_B$). However, the states resolved in $\psi_A$ and $\psi_B$ by measurements on the separated subsystems derive from elements of the coset of $\Psi$ in $\mathcal{F}$. These elements can be distinguished when we lift back from the quotient space in which they are defined to be equivalent.

For example, consider an entangled state in the tensor product space $\mathcal{H}_A \otimes \mathcal{H}_B$ with the structure, up to an arbitrary overall phase,
\begin{equation}
	\Psi = v_A \otimes v_B + r(w_A \otimes w_B) ,
\end{equation}
where $v_A, w_A \in \mathcal{H}_A$,  $v_B, w_B \in \mathcal{H}_B$, and $r \in \mathbb{C}$. First note the obvious identification that $\Psi$ corresponds to an equivalence class of a superposition of points in $\mathcal{F}(U_A \times V_B)$:
\begin{align}
	(v_A \otimes v_B) &+ r(w_A \otimes w_B) \nonumber \\
	&= [(v_A, v_B) + r(w_A, w_B)]  \\
	&= [(v_A, v_B)] + [r(w_A, w_B)] , \nonumber
\end{align}
where $[x]$ means equivalence class, that is, the coset of $x$ in the quotient space. Now it becomes clear how the single wavefunction in $\mathcal{T}$ has different representatives in $\mathcal{F}$; notice that the coset $[r(w_A, w_B)]$, for example, has representatives $(rw_A, w_B)$ and $(w_A, rw_B)$. It is in this sense that the wavefunction has statistical properties because each of these representatives is statistically equally likely to be probed by measurements on the separated subsystems, as we develop below. For the entangled states, $r$ is the phase of the superposition (for the Bell states, it will be $1$ or $-1$), which is what becomes important for measurements. We then say that $(rw_A, w_B)$ and $(w_A, rw_B)$ denote possible `contextual phases'. 

\begin{figure}[h]
	\centering
	\includegraphics[width=0.8\textwidth]{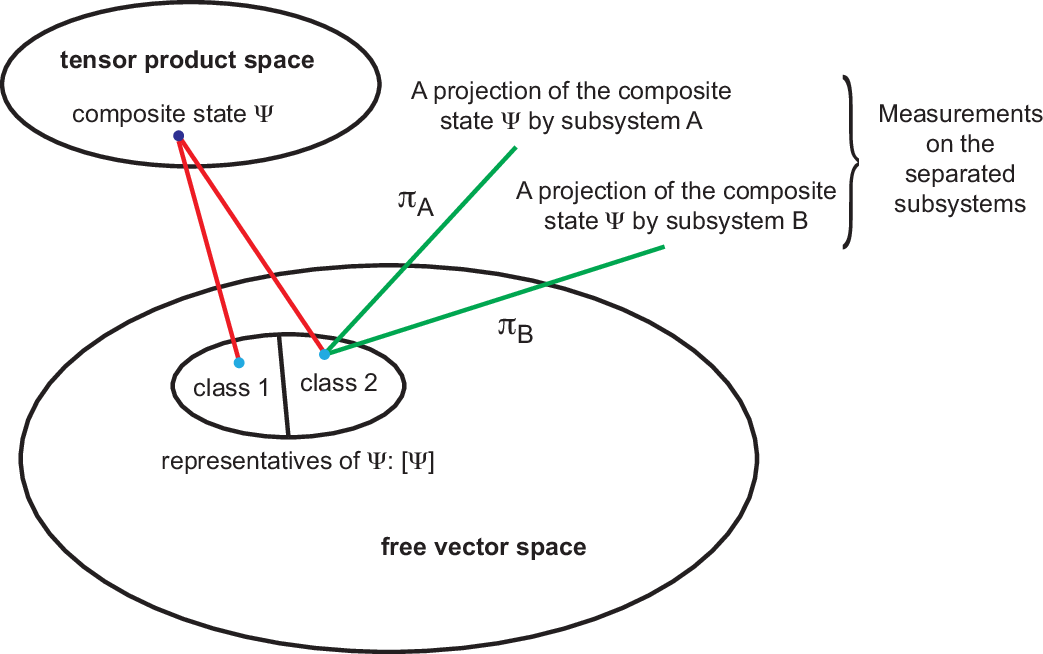}
	\caption{Diagram showing how a composite state in the tensor product space $\mathcal{T}$ is associated to a coset in the free vector space $\mathcal{F}$  comprising class 1 and class 2 representatives. These representatives can, in turn, be associated to vectors in the Hilbert spaces for the separated subsystems. }\label{fig2}
\end{figure}

The contextual phase is simply the phase of the superposition in $\Psi$; it is nothing more than that. When we associate $\Psi$ to representative points in the free vector space $\mathcal{F}(U_A \times V_B)$, this phase needs to be explicitly associated with points in $U_A$ or $V_B$, and hence one of the vectors in $\mathcal{H}_A$ or $\mathcal{H}_B$ (we use the example of an antisymmetric state here). When the subsystems are separated, the phase of the superposition ends up associated in various ways with each subsystem state. These two possibilities---which are equally likely---are denoted class 1 and class 2. The label 1 or 2 is arbitrary. The goal is to sort the sets of points in $\mathcal{F}(U_A \times V_B)$ by the measurement outcomes in $\psi_A$ and $\psi_B$. A summary of the relationships between the relevant spaces and contextual phase classes is given in Fig. 2.

The local phases $\phi_A$ and $\phi_B$ for subsystem states in $\mathcal{H}_A$ and $\mathcal{H}_B$, respectively, can take various values\cite{sepsubs}. Nevertheless, we group all the representatives into two classes that decide opposite and definite measurement outcomes. We only need to consider a specific example of $\Psi$ from each class rather than detailing all possible forms of $\Psi$ in the free vector space. Note that in the examples we have set one of these phases to be 0 because we are focusing on the maximally entangled states and measurements in the $x-$ or $z-$basis. However, if we set these phases to be appropriate multiples of $\frac{\pi}{2}$, then we can account for measurements in the $y-$basis. Refer to the quantum beam splitter example in Ref. \cite{collapse}.

\section{Additional example}

Consider the following elements of $\mathcal{F}(U \times V)$:
\begin{eqnarray*}
	v = (x_1, y) + (x_2,y) \\
	v' = (x_1 + x_2, y).
\end{eqnarray*}
The $v$ and $v'$ are distinct elements of the free vector space, but by quotienting out the subspace of bilinear relations to form the tensor product space $\mathcal{T}(U \otimes V)$, we define $v$ and $v'$ to be equivalent. 

Now let's consider the projection of each of $v$ and $v'$ from $\mathcal{F}(U \times V)$ to $U$. This map extracts the $U$-components such that for any formal sum $\sum_i \lambda_i (x_i, y_i)$ (Eq. 2),
\begin{equation*}
	\pi_U \Big( \sum_i \lambda_i (x_i, y_i) \Big) = \sum_i \lambda_i x_i .
\end{equation*}
Applying this projection to $v$ and $v'$ we see that 
\begin{equation*}
	\pi_U(v) = \pi_U(v') = x_1 + x_2
\end{equation*}
because in the standard Banach or Hilbert space $U$, $x_1 + x_2$ is a single vector. 

This simple example emphasizes that the free vector space is very large and it differentiates vectors in subtle ways. The projections lose much of this information because the spaces $U$ and $V$ are small in comparison. The quotient space $U \otimes V$ is a kind of middle ground.

\section{Measurements on separated subsystems}
Here we bring together concepts developed in the preceding sections to show how measurements of entangled states, via their separated subsystems, yields definite measurement outcomes in practice.

\subsection{The statistical basis for measurement outcomes}
At the heart of the EPR paper\cite{EPR} is the concept of measurements on subsystems of an entangled state, where the subsystems are separated so that measurements can be performed on each subsystem independently. That allows correlations to be revealed in the measurement outcomes. Such measurements are possible because quantum subsystems include electrons, atoms, photons, and so on---particles that can be entangled and subsequently physically separated in space. Similarly, in the measurement devices described in Sec. 2, the readout state is entangled among the output `ports'---and these ports are the separated subsystems. 

What is clear about measurements on separated subsystems is that we can predict the average results using the density matrix $\rho$ for any entangled state $\Psi \in U \otimes V$. By tracing out one subsystem, we can predict the probabilistic results for the other subsystem from the resulting mixed state in the subsystem of interest. As described earlier in this paper, there is not a way to do something analogous for the state vector $\Psi$ (see also Ref. \cite{sepsubs}). Nor does the analysis in terms of density matrices for the subsystems tell us anything about correlations among joint measurements on the two subsystems. 

We have established that the separated subsystems of $\Psi$ can be analyzed by a suitable projection map that defines the states of the separated subsystems. We have further shown how this reveals hidden phase information (that is irrelevant for the composite states in $U \otimes V$). This hidden phase information is the basis for definite measurement outcomes without the need for an \emph{ad hoc} notion of reduction of the state vector because it controls constructive and destructive interference among the measurement outcomes. We can now unravel the probabilistic determinations given by the density matrices for the subsystems to obtain the distribution of \emph{joint} definite measurement outcomes that must underlie the probabilistic prediction. Of course, the Copenhagen interpretation proposes a partial solution based on random collapse of the wavefunction to yield a definite measurement outcome, but it fails to explain why the collapse of the states projected to the separated subsystems are correlated. 

Our claim is that the overall solution is hidden by the quotient map, so we cannot obtain the required insight by working with the vectors in the tensor product space. Instead, we need to look more generally at their representatives in the overarching  free vector space, Fig. 2. Then we find the reason not only for why wavefunctions give definite measurement outcomes when experiments probe the appropriate separated subsystems, but also we see why joint measurements are correlated and how measurement statistics arise even though every wavefunction in the sequence of measurements is identical in the tensor product space. The key statement regarding this latter point is the following.

While we are not suggesting that there is an ensemble of wavevunctions in the usual statistical sense, an entangled state is a source of statistical \emph{measurement outcomes}. Given $\Psi \in \mathcal{T}(U_A \otimes V_B)$, we have a coset $[\Psi] \in \mathcal{F}(U_A \times V_B)$ that comprises the sets $\{\text{class 1}\}$  and $\{\text{class 2}\}$, differentiated by contextual phases. Whether a measurement reveals an element from $\{\text{class 1}\}$  or from $\{\text{class 2}\}$ is equally likely. For example, consider the elements $(r v, w) - r (v,w)$ and $(v, r w) -  r (v,w)$ from $\mathcal{R}$. In the quotient space, these elements mean that $[r(v, w)] $ has representatives of $(rv, w)$ and $(v, rw)$, with equal statistical likelihood. We can insert $r  = e^{i\phi}$, making clear how the contextual phase, which becomes explicitly associated with $\psi_A$ and/or $\psi_B$, derives from $\Psi$, wherein it serves as the phase of the superposition.

With suitable physical restrictions in place, there are still several equivalent elements of $\mathcal{F}(U_A \times V_B)$ that represent an entangled state in $U_A \otimes V_B$. Up to an arbitrary global phase, there are eight obvious ways to incorporate contextual phases in each $\Psi$. These representations of $\Psi$, which are all equivalent in $U_A \otimes V_B$, can be grouped into two distinct classes, labelled 1 and 2, distinguished by the way contextual phase is associated with the subsystems. Then we only need to take one representative instance of contextual phase from each class. Class 1 elements of $\mathcal{F}$ display complementary measurement outcomes for the separated subsystems compared to class 2 elements of $\mathcal{F}$. For example, for an antisymmetric entangled state we have possible points listed in Table 1, where $\phi$ and $\phi'$ equal $(2n+1)\pi$. Any of the $\Psi$ are identical in $U_A \otimes V_B$, but can be distinguished as class 1 or class 2 when measurements are performed on the separated subsystems. 

\begin{table}[!h]
	\caption{Examples of class 1 and 2 states in $\mathcal{F}$ for an antisymmetric entangled state.}
	\label{table_1}
	\begin{tabular}{ll}
		\hline
		Class 1 & Class 2  \\
		\hline
		$(v_A, v_B) + (w_A, e^{i\phi}w_B)$ &  $(v_A, v_B) + (e^{i\phi}w_A, w_B)$   \\
		$(v_A, e^{i\phi'}v_B) + (w_A, w_B)$ & $(e^{i\phi'}v_A, v_B) + (w_A, w_B)$ \\
		$(e^{i\phi'}v_A, e^{i\phi'}v_B) + (e^{i\phi}w_A, w_B)$ & $(e^{i\phi'}v_A, e^{i\phi'}v_B) + (w_A, e^{i\phi}w_B)$  \\
		$(e^{i\phi'}v_A, v_B) + (e^{i\phi}w_A, e^{i\phi}w_B)$ & $(v_A, e^{i\phi'}v_B) + (e^{i\phi}w_A, e^{i\phi}w_B)$  \\
		\hline
	\end{tabular}
	\vspace*{-4pt}
\end{table}

\subsection{Discussion of measurement context and basis}
In prior work, Auff\'eves and Grangier\cite{CSM}, and Grangier\cite{Grangier2021, hidden9}, have developed an ontological framework enabling quantum mechanical measurements to be better understood with regard to physical realism. The framework is called `contexts, systems, and modalities' (CSM). The way measurements of quantum states are described using contextual phases closely parallels the CSM ontology---hence the label \emph{contextual} phase. The two theories are complementary, and it will be interesting to develop the parallels more deeply in future work. CSM is a top-down theory motivated by an overarching viewpoint that reasons how measurement outcomes depend on the \emph{combination} of the state and all the factors prescribed by the measurement system, termed the contexts. It is shown below how this is precisely the way measurement works for the contextual phase theory. The difference is that contextual phase theory is developed from the bottom-up.

For the description of measurements using contextual phase theory, we require that, as indicated in the postulates, when a measurement is performed, only one result is obtained. Two sequential identical measurements must give the same outcomes.  The point we will focus on here is to explain how identical wavefunctions can be presented in different ways to expose different possible underlying superpositions. These different representations of the \emph{same} wavefunction encode the concept of measurement context together with the context provided by contextual phase. 

A way to grasp physically why equivalent representations of $\Psi$ give the measurement context, is to recognize that for any measurement conditions, the superpositions of the separated subsystems must be able to interfere in such a way to guarantee that only one measurement result is obtained. To accomplish this, we view the wavefunction of the composite system in different ways (bases) compatible with the measurement to be performed. Then, these states, expressed in the pre-image form to accommodate the association of these states with those projected to the Hilbert spaces of the separated subsystems \emph{carry information both about the state and about the measurement context}, just as postulated in CSM theory. An example is $x-$basis versus $z-$basis measurement conditions. In Sec. 8 it will become clear why there is an intimate relationship between basis and contextual phase.

Each entangled wavefunction $\Psi$ can be written in many equivalent ways. The Hadamard gate $H = \frac{1}{\sqrt{2}} \bigl[\begin{smallmatrix} 1 & 1 \\ 1 & -1 \end{smallmatrix}\bigr]$ transforms each subsystem from the $x$-basis to the $z$-basis, so the transformation of the composite state $\Psi$ is carried out using $H \otimes H$. Alternatively, we can write out the $x$-basis form explicitly and simplify it to the $z$-basis by collecting terms (see below). Also note that usually the $x$-basis form of $\Phi_-$ is labeled $\Psi_+$ and, conversely, the $x$-basis form of $\Psi_+$ is labeled $\Phi_-$. Here we do not change the label. Instead we keep the $z$-basis notation regardless of the basis in order to emphasize that these wavefunctions are identical.

For example, $\Psi_+$ is given in standard ($z-$basis) form by
\begin{equation}
	\Psi_+ = \frac{1}{\sqrt{2}} \Big[ |0\rangle_A \otimes |1\rangle_B + |1\rangle_A \otimes |0\rangle_B  \Big] .
\end{equation}
Using the notation $|0'\rangle \equiv \frac{1}{\sqrt{2}}(|0\rangle + |1\rangle)$, $|1'\rangle \equiv \frac{1}{\sqrt{2}}(|0\rangle - |1\rangle)$, and also $|0''\rangle \equiv \frac{1}{\sqrt{2}}(|0\rangle + i|1\rangle)$, $|1''\rangle \equiv \frac{1}{\sqrt{2}}(|0\rangle - i|1\rangle)$, we can write $\Psi_+$ in various other forms that include the $y-$basis and $x-$basis representations:
\begin{eqnarray}
	\Psi_+ = \frac{1}{\sqrt{2}} \Big[ |0''\rangle_A \otimes |0''\rangle_B - |1''\rangle_A \otimes |1''\rangle_B  \Big] , \\
	\Psi_+ = \frac{1}{\sqrt{2}} \Big[ |0'\rangle_A \otimes |0'\rangle_B - |1'\rangle_A \otimes |1'\rangle_B  \Big] .
\end{eqnarray}
The equivalence is seen by expanding the $x-$basis representation as follows:
\begin{align*}
	\Psi_+ &= \frac{1}{\sqrt{2}} \Big[ \frac{1}{2} ( |0\rangle_A|0\rangle_B + |0\rangle_A|1\rangle_B + |1\rangle_A|0\rangle_B + |1\rangle_A|1\rangle_B \\
	&- |0\rangle_A|0\rangle_B + |0\rangle_A|1\rangle_B + |1\rangle_A|0\rangle_B - |1\rangle_A|1\rangle_B ) \\
	&= \frac{1}{\sqrt{2}} \Big[ |0\rangle_A \otimes |1\rangle_B + |1\rangle_A \otimes |0\rangle_B  \Big] .
\end{align*}
Expanding the $y-$basis expression gives the same result up to an overall phase. 

These representations, Eqs. 7--9, of the wavefunction are all identical, but they are composed explicitly of different superpositions. For example, the $z-$basis form is the obvious and simplest superposition of product functions. Whereas, the $x-$basis expression superposes (in the computational basis) products so that $|0\rangle_A |1\rangle_B + |1\rangle_A |0\rangle_B$ interferes constructively with itself, while $|0\rangle_A|0\rangle_ B + |1\rangle_A|1\rangle_B$ interferes destructively with itself. This makes no difference to the vectors in $\mathcal{H}_A \otimes \mathcal{H}_B$ and observables  in this space. But, the different representations of $\Psi_+$ give distinct superpositions of states for the separated subsystems, in $\mathcal{H}_A$ and $\mathcal{H}_B$. Hence the measurement outcomes for the separated subsystems vary according to measurement conditions. 

Consider the example of $\Psi_+$. Joint measurement outcomes for the operator $\sigma_z$ on each of the separated subsystems gives anticorrelated results $0$ and $1$ (the operator acts on our $x-$basis representation of $\Psi_+$, Eq. 9, in order to give an observable outcome). Below we give explicit details about how this works. Whereas the operator $\sigma_x$ on each of the separated subsystems gives perfectly correlated results $0'$ and $0'$ or $1'$ and $1'$ (here the operator acts on our $z-$basis representation of $\Psi_+$, Eq. 7, in order to give an observable outcome). It is through the way the different representations of $\Psi$ are conveyed to the states of the separated subsystems that endow the state with measurement context, consistent with the CSM theory.

\subsection{Definite measurement outcomes}
Equipped with the contextual phase, an appropriate representation of $\Psi$ yields state vectors for the separated subsystems that collapse when measured. `Collapse' in the context means the vectors simplify by interference. This ensures that only one result is obtained from the measurement of each subsystem, as specified in the postulates. Notice that there is no need to invoke the usual \emph{ad hoc} notion of collapse here. Collapse in the setting of contextual phase theory corresponds to the following: The superposition in $\mathcal{F}(U_A \times V_B)$ implies an associated superposition of vectors in each of $\psi_A$ and $\psi_B$. When we measure either of these \emph{separated states}, the superposition collapses locally, in the sense that the vector in the local Hilbert space naturally simplifies to enable a definite measurement outcome. That is, the local superposition is set up---by the construction of the entangled composite state and its measurement context---so that it gives a single definite observable. 

By this reasoning, contextual phase theory gives a plausible mechanism for explaining the concept of collapse (or reduction of the wavepacket) that is implied by the quantum measurement postulates. 

Here we give specific examples of measurements in the $z-$ and $x-$ basis. A summary of the process is given in Fig. 3.  Consider $\Psi_+$ presented in the following basis form:
\begin{equation*}
	\Psi_+ = \frac{1}{\sqrt{2}} \Big[ |0'\rangle_A \otimes |0'\rangle_B - |1'\rangle_A \otimes |1'\rangle_B  \Big] .
\end{equation*}
We can re-write this informally as a pre-image, with a class 1 contextual phase choice, to foreshadow the class 1 representation in $\mathcal{F}$, but retain the familiar tensor product notation:
\begin{equation}
	\Psi_+ = \frac{1}{\sqrt{2}} \Big[ |0'\rangle_A \otimes |0'\rangle_B + |1'\rangle_A \otimes |-1'\rangle_B  \Big] ,
\end{equation}
which is given in terms of the generators for the free vector space, written without normalization, as
\begin{equation}
	\Psi_+ = [ (e_0 + e_1, f_0 + f_1) + (e_0 - e_1, -f_0 + f_1)  ].
\end{equation}

A free vector space is formed from a set, where the elements of the set are called generators. Here, $\{ e_0, e_1 \}$ are the generators for subsystem A and $\{ f_0, f_1 \}$ are the generators for subsystem B. Now, we have a one-to-one map from these generators to the usual basis states in the relevant Hilbert spaces. For example we can equate $e_0 \rightarrow |0\rangle$ and $e_1 \rightarrow |1\rangle $. 

The canonical map on the corresponding formal linear combination in $\mathcal{F}(U_A \times V_B)$ gives the formal linear combination in $\mathcal{F}(U_A)$:
\begin{equation}
	u_A = \frac{1}{2}  [ (e_0 + e_1) + (e_0 - e_1)  ].
\end{equation}

Notice that this has been written in a way that emphasizes the superposition of possible measurement outcomes. That superposition must be preserved for the subsystems to remain entangled and it is the reason to expect violation of Bell inequalities. The point is that this vector does not simplify or collapse simply because the subsystems are separated. If that happened, then the entanglement would be destroyed and we are left with a classical state. We know this must be the case because entanglement between widely separated systems, such as photons, is maintained. 

When a measurement is carried out on subsystem A, Eq. 12  maps to the vector in $\mathcal{H}_A$
\begin{align*}
	\psi_A &= \frac{1}{2}  \underbrace{ ( |0\rangle + |1\rangle + |0\rangle - |1\rangle  ) }_{ \text{simplifies when measured} }  \\
	&= |0\rangle .
\end{align*}
Notice that the vector $\psi_A$ simplifies (collapses) when a measurement is made on the separated subsystem to give a definite outcome based on the expectation value for the operator acting on $|0\rangle$.

This simplification, or collapse of this projected state only happens when a measurement is performed. What is it that makes the vector `collapse' when it is measured? This phenomenon must be caused by the operation of the measuring device itself. A plausible hypothesis comes from considering a prototypical polarizing beam splitter (Fig. 1b). Let's say that Eq. 10 refers to an entangled polarization state of a photon pair, where $|0\rangle$ means the horizontal (H) polarization basis state and $|1\rangle$  the vertical (V) polarization basis state. The H-polarized component is reflected in the device, while the V-polarized component is transmitted. That allows an interpretation of why the class 1 $\psi_A$ gives the definite outcome of reflecting a single H-polarized photon. The H-polarized components of $\psi_A$ add constructively at the output port 2, whereas the V-polarized components interfere destructively at port 3. We can interpret this as a self-interference phenomenon in the measuring device that emphasizes the wave-particle duality of quantum systems.  

Similarly for a measurement on subsystem B, in $\mathcal{H}_B$, where the canonical map on the corresponding formal linear combination in $\mathcal{F}(U_A \times V_B)$ gives the formal linear combination in $\mathcal{F}(V_B)$:
\begin{equation}
	v_B = \frac{1}{2}  [ (e_0 + e_1) - (e_0 + e_1)  ],
\end{equation}
which maps to the vector in $\mathcal{H}_B$
\begin{align*}
	\psi_B &= \frac{1}{2}  \underbrace{ ( |0\rangle + |1\rangle - |0\rangle + |1\rangle  ) }_{ \text{simplifies when measured} }  \\
	&= |1\rangle .
\end{align*}

What we mean by collapse of the local vector in these equations is that, for some operator $T$,  $\langle \psi_A | T | \psi_A \rangle = \langle 0 | T | 0 \rangle$  and $\langle \psi_B | T | \psi_B \rangle = \langle 1 | T | 1 \rangle$. Working through the case of $\Psi_+(\text{class 2})$, we obtain the opposite outcomes, that is, $\langle \psi_A | T | \psi_A \rangle = \langle 1 | T | 1 \rangle$  and $\langle \psi_B | T | \psi_B \rangle = \langle 0 | T | 0 \rangle$. Therefore, the $x-$basis presentation of $\Psi$ enables predictions of the $z-$basis measurement outcomes.

\begin{figure}[h]
	\centering
	\includegraphics[width=0.8\textwidth]{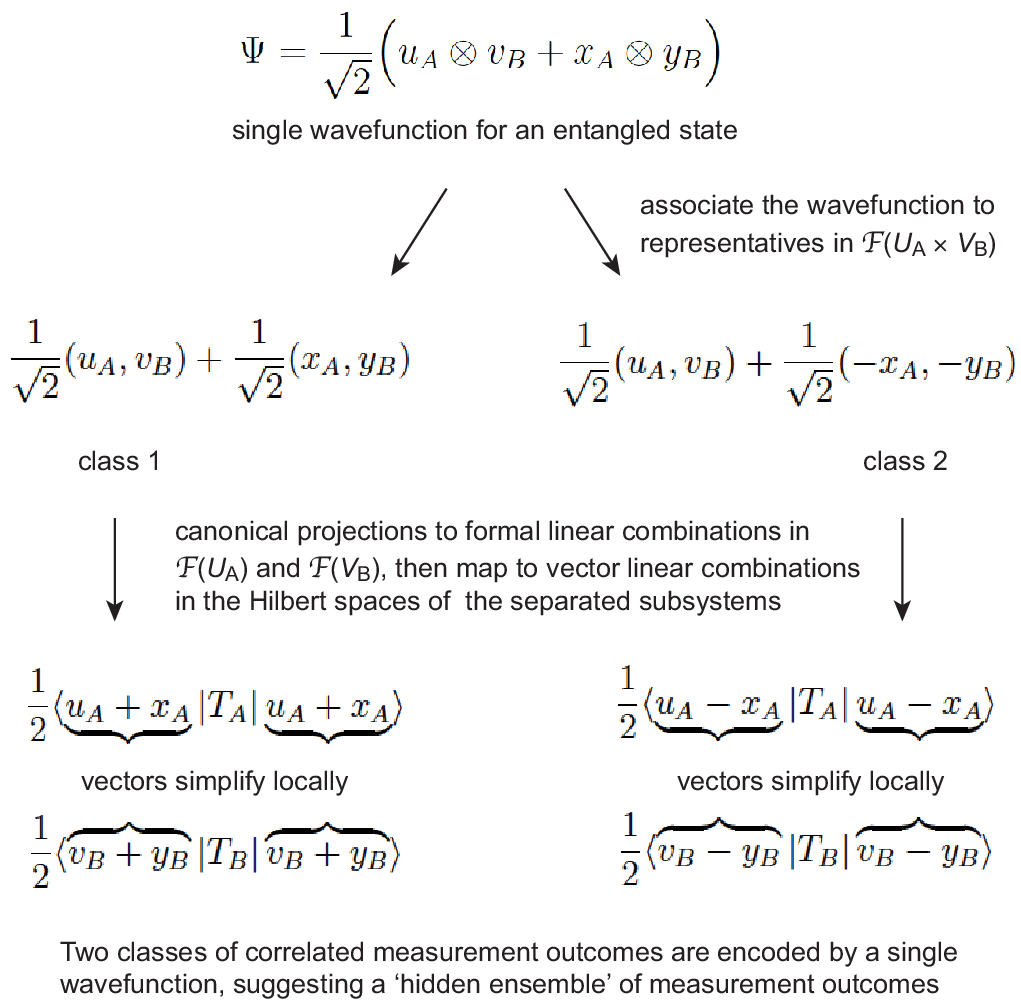}
	\caption{Summary scheme showing how the class 1 and class 2 representatives in the free vector space $\mathcal{F}$ of a vector in the tensor product space $\mathcal{T}$ are associated to vectors in the Hilbert spaces for the separated subsystems. }\label{fig3}
\end{figure}

Compare those results to $\Psi_+$ presented in the equivalent form:
\begin{equation*}
	\Psi_+ = \frac{1}{\sqrt{2}} \Big[ |0\rangle_A \otimes |1\rangle_B + |1\rangle_A \otimes |0\rangle_B  \Big] .
\end{equation*}
We can re-write this informally as a pre-image, with a class 1 contextual phase choice, to foreshadow the class 1 representation in $\mathcal{F}$, but retain the familiar tensor product notation:
\begin{equation}
	\Psi_+ = \frac{1}{\sqrt{2}} \Big[ |0\rangle_A \otimes |1\rangle_B + |1\rangle_A \otimes |0\rangle_B  \Big] ,
\end{equation}
or
\begin{equation}
	\Psi_+ = \frac{1}{\sqrt{2}} \Big[ |0\rangle_A \otimes |1\rangle_B + |-1\rangle_A \otimes |-0\rangle_B  \Big] .
\end{equation}
Let's consider the class 1 representation in Eq. 15. Then the canonical map on the corresponding formal linear combination in $\mathcal{F}(U_A \times V_B)$ gives the formal linear combination in $\mathcal{F}(U_A)$:
\begin{equation}
	u_A = \frac{1}{2}  ( e_0 + e_1 ),
\end{equation}
which maps to the vector in $\mathcal{H}_A$
\begin{align*}
	\psi_A &= \frac{1}{2}  \underbrace{ ( |0\rangle + |1\rangle  }_{ \text{simplifies} }  \\
	&= |0'\rangle .
\end{align*}
Notice that the vector $\psi_A$ simplifies when a measurement is made on the separated subsystem to give a definite outcome based on the expectation value for the operator acting on $|0'\rangle$. In this case of $x-$axis detection, the measurement outcomes are perfectly correlated, so we obtain the same result for subsystem B.

What selects the representative that produces a single outcome? The precise representative selected during any measurement is random because we know that a sequence of measurements of one subsystem yields a random sequence of measurement outcomes. Why this should be the case will become more apparent in Sec. 8. The experimenter can set the basis for the measurement. However, each basis choice comes with two associated contextual phase class choices, which cannot be set or known by the experimenter. Thus, for each interaction between a wavefunction and measuring device, the contextual phase class is randomly selected.  

How do the measurements work when one subsystem (say B) is destroyed before measurements are made on A? Using the example of $\Psi_+$, if the contextual phases for the entangled system are configured as class 1, then measurement outcomes for the isolated $\psi_A$ will be expectation values of $|0\rangle$ or $|0'\rangle$, while for class 2 phases, the outcomes will be expectation values of $|1\rangle$ or $|1'\rangle$. In either case, we have a good basis for the isolated system and regardless how we set the measuring device (e.g. polarizer set to $H/V$ or $H'/V'$), then a sequence of measurements yields the appropriate random sequence of outcomes. This example gives an important consistency check for the states relevant to the separated subsystems---they are appropriate for when the system remains entangled with the other subsystem, or for the case where the second subsystem is destroyed and there is only one relevant Hilbert space for the states ($\mathcal{H}_A$).

The set of bipartite entangled states are documented in the informal `pre-image' form in Table 2. Measurement outcomes are collected in Table 1 of Ref. \cite{collapse}.

\begin{table}[!h]
	\caption{`Pre-images' (using informal notation) of bipartite entangled states showing the origin of class 1 and 2 measurement outcomes.}
	\label{table_2}
	\begin{footnotesize}
		\begin{tabular}{ll}
			\hline
			$x-$basis\textsuperscript{a} & $z-$basis  \\
			\hline
			$\Psi_-(\text{class 1}) = \frac{1}{\sqrt{2}} \Big[ |1'\rangle_A \otimes |0'\rangle_B + |0'\rangle_A \otimes |-1'\rangle_B  \Big] $ &  $\Psi_-(\text{class 1}) = \frac{1}{\sqrt{2}} \Big[ |0\rangle_A \otimes |1\rangle_B + |-1\rangle_A \otimes |0\rangle_B  \Big]$  \\
			$\Psi_-(\text{class 2}) = \frac{1}{\sqrt{2}} \Big[ |1'\rangle_A \otimes |0'\rangle_B + |-0'\rangle_A \otimes |1'\rangle_B \Big]$ & $\Psi_-(\text{class 2}) = \frac{1}{\sqrt{2}} \Big[ |0\rangle_A \otimes |1\rangle_B + |1\rangle_A \otimes |-0\rangle_B  \Big] $ \\
			$\Phi_-(\text{class 1}) = \frac{1}{\sqrt{2}} \Big[ |1'\rangle_A \otimes |0'\rangle_B + |0'\rangle_A \otimes |1'\rangle_B \Big]$ & $\Phi_-(\text{class 1}) = \frac{1}{\sqrt{2}} \Big[ |0\rangle_A \otimes |0\rangle_B + |-1\rangle_A \otimes |1\rangle_B  \Big]$ \\
			$\Phi_-(\text{class 2}) = \frac{1}{\sqrt{2}} \Big[ |1'\rangle_A \otimes |0'\rangle_B + |-0'\rangle_A \otimes |-1'\rangle_B  \Big]$ & $\Phi_-(\text{class 2}) = \frac{1}{\sqrt{2}} \Big[ |0\rangle_A \otimes |0\rangle_B + |1\rangle_A \otimes |-1\rangle_B  \Big] $ \\
			$\Phi_+(\text{class 1}) = \frac{1}{\sqrt{2}} \Big[ |0'\rangle_A \otimes |0'\rangle_B + |1'\rangle_A \otimes |1'\rangle_B  \Big]$ &  $\Phi_+(\text{class 1}) = \frac{1}{\sqrt{2}} \Big[ |0\rangle_A \otimes |0\rangle_B + |1\rangle_A \otimes |1\rangle_B  \Big]$  \\
			$\Phi_+(\text{class 2}) = \frac{1}{\sqrt{2}} \Big[ |0'\rangle_A \otimes |0'\rangle_B + |-1'\rangle_A \otimes |-1'\rangle_B  \Big]$ & $\Phi_+(\text{class 2}) = \frac{1}{\sqrt{2}} \Big[ |0\rangle_A \otimes |0\rangle_B + |-1\rangle_A \otimes |-1\rangle_B  \Big]$ \\
			$\Psi_+(\text{class 1}) = \frac{1}{\sqrt{2}} \Big[ |0'\rangle_A \otimes |0'\rangle_B + |1'\rangle_A \otimes |-1'\rangle_B  \Big] $ & $\Psi_+(\text{class 1}) = \frac{1}{\sqrt{2}} \Big[ |0\rangle_A \otimes |1\rangle_B + |1\rangle_A \otimes |0\rangle_B  \Big]$ \\
			$\Psi_+(\text{class 2}) = \frac{1}{\sqrt{2}} \Big[ |0'\rangle_A \otimes |0'\rangle_B + |-1'\rangle_A \otimes |1'\rangle_B  \Big]$ & $\Psi_+(\text{class 2}) = \frac{1}{\sqrt{2}} \Big[ |0\rangle_A \otimes |1\rangle_B + |-1\rangle_A \otimes |-0\rangle_B  \Big]$ \\
			\hline
		\end{tabular}
		\tabnote{\textsuperscript{a} Usually the $x$-basis form of $\Phi_-$ is labeled $\Psi_+$ and, conversely, the $x$-basis form of $\Psi_+$ is labeled $\Phi_-$. Here we do not change the label. Instead we keep the $z$-basis notation regardless of the basis in order to emphasize that these wavefunctions are identical.}
	\end{footnotesize}
	\vspace*{-4pt}
\end{table}

\section{Nonlocality}

Nonlocality lies at the heart of the EPR paradox\cite{Kupczynski2006}. The concept means that when we perform measurements on two subsystems (A and B) that are far apart and equally likely to yield either of two measurement outcomes, and we accept that measurement of one subsystem does not disturb the other, the outcomes of these measurements are correlated. As stated by Vaidman\cite{Vaidman2019}: ``Given entangled particles placed at a distance, a measurement on one of the particles instantaneously changes the quantum state of the other, from a density matrix to a pure state''. In this section we focus on how contextual phase classes not only ensure measurements on each subsystem give definite outcomes in random sequence, but they also enforce correlations \emph{between} measurement outcomes on the subsystems. 

Kupczynski nicely summarizes three possible ways of rationalizing correlations of the entangled pair of subsystems\cite{Kupczynski2006}:
\begin{enumerate}
	\item Assume that all the pairs of subsystems are the same, and that the random outcomes of measurements on each subsystem are correlated by some spooky action at a distance.
	\item Assume a statistical ensemble of different pairs of subsystems, and for each pair the results of all measurements are predetermined and correlated in the moment of production.
	\item Assume that the long-range correlations of the subsystem measurement results are due to their particular preparation by a common source, but at the same time, assume that the result of any single measurement is not predetermined.
\end{enumerate}

But contextual phase theory suggests a fourth possibility:
\begin{enumerate}
	\setcounter{enumi}{3}
	\item Assume that all composite states are the same, and that there is a statistical distribution of \emph{subsystem} pairs owing to the contextual phases, which, in turn, implies that the correlation between measurement pairs is statistically predetermined. 
\end{enumerate}

This proposal does not imply hidden local variables, since the correlations arise solely from the quantum state and its measurement context. The contextual phase comes from the phase already present in the superposition. The way a single wavefunction in the tensor product basis is associated to multiple classes of measurable separated states in the free vector space representation may be the kind of ensemble view of a wavefunction hoped for by Einstein\cite{FineBook, Werner2014}. 

The correlation between measurement outcomes depends on the measurement conditions, so it is not possible to realize simultaneously all possible outcomes, as is well known\cite{Mermin1993}. For example, $\Psi_+$ when both subsystems are measured in the $z-$basis will show anti-correlated outcomes. But measured in the $x-$basis the results are correlated. See Table 1 of ref \cite{collapse}. The correlations are already encoded in the entangled states. These correlations are translated to $\psi_A$ and $\psi_B$, and hence to the classical measurement outcomes.

\section{Measurement basis, contextual phases, and nonlocal correlations}
In Appendix B we show an exploratory example of measurements on entangled states that emphasizes the construction of entangled states from separable states. The approach exposes a connection between contextual phase classes in one basis and measurements in an orthogonal basis. Here we give a precise example that shows how basis rotation in the free vector space generates both basis representations of the wavefunction and associated contextual phase classes.  We thereby obtain an understanding for the relationship between superpositions of states with different contextual phase class and measurement basis.

In Appendix B an analysis is documented that shows fundamentally how entangled states, together with measurement basis and contextual phase classes, can be built from a general class of separable states. Here we focus on a more concise, universal way of showing the close relationship between contextual phase classes and measurement bases. Then we will show that contextual phase classes, like measurement basis, affect measurement outcomes, but do not control the joint measurement correlations. That is, they are not hidden local variables.

Each basis vector is given in terms of a normalized state of a two-level system, 
\begin{equation}
	u = \alpha |0\rangle + \beta |1\rangle,
\end{equation}
where $\alpha, \beta \in \mathbb{C}$ and $|\alpha|^2 + |\beta|^2 = 1$. The vector $u$ can be identified with a point lying on the circle $\mathbb{S}^3$ embedded in $\mathbb{C}^2 \cong \mathbb{R}^4$. An entangled state
\begin{equation*}
\Psi = \frac{1}{\sqrt{2}} (u_A \otimes u_B \pm w_A \otimes w_B)
\end{equation*}
is a linear combination of points, each point lying on the six-dimensional manifold $\mathbb{S}_A^3 \times \mathbb{S}_B^3$. Usually we can greatly simplify this analysis because states $u$ that differ by a global phase $e^{i\phi}$ are indistinguishable, we map all points in $\mathbb{S}^3$ that differ only by global phase to one point each on $\mathbb{S}^2$ by the Hopf fibration $\mathbb{S}^3/\mathbb{S}^1$, giving a surface of equivalence points known as the Bloch sphere. The main message of the present work, however, is that to understand what happens to the phase factor \emph{within} the superposition when measurements are performed on the separated subsystems, we should retain phase information and only factor out the global phase from the superposition state (the entangled state). 

To be more explicit, what we can show is that, when we retain the full phase content of the state, the usual rotation of the basis generates different basis representations of the states and contextual phase classes--suggesting their close relationship. To perform general basis rotations we can define three Euler angles to define all possible transformations\cite{SimonMukunda}, or we can represent each pair of basis vectors as a unit quaternion\cite{NaiveLie} and perform rotations explicitly in terms of points on $\mathbb{S}^3$. It is sufficient, however, to consider the case when the coefficients $\alpha$ and $\beta$ are real. The basis vectors are written in vector form as $|0\rangle = 
\begin{pmatrix}
	1 \\
	0  
\end{pmatrix} $
and $|1\rangle = 
\begin{pmatrix}
	0 \\
	1  
\end{pmatrix} $. 

We can now apply rotations using the usual rotation matrix:
\begin{equation}
	R_{\theta} = 
	\begin{pmatrix}
		\cos \theta & -\sin \theta \\
		\sin \theta & \cos \theta 
	\end{pmatrix}.
\end{equation}
We thus define the operator $U = R_{\zeta} \otimes R_{\theta}$ such that 
\begin{equation}
	U(u \otimes w) = (R_{\zeta} u) \otimes (R_{\theta} w).
\end{equation}

Now consider joint measurements on various common bases. Set $\zeta = \theta = \frac{\pi}{4}$.Applying the same rotation to both subsystems, i.e. $U = R_{\theta} \otimes R_{\theta}$,  leaves the state unchanged. Only the basis in which  we determine or measure the wavefunction changes.  Because the coefficients are real, we  can visualize them  on $\mathbb{S}^1_A \times \mathbb{S}^1_B$. A sequence of $\frac{\pi}{4}$ rotations generates a cycle of states and contextual phase classes. For instance, consider $\Psi_-$. Start by writing down the state in the $z$-basis and then apply $U$ sequentially to obtain

\begin{tikzpicture}[node distance = 3cm, auto, >= latex]
	\node[font=\LARGE] (A) at (90 : 3cm) {$\mathcolor{red}{\Psi^{(1,z)}_-}$};
	\node[font=\LARGE] (B) at (45 : 3cm) {$\mathcolor{blue}{\Psi^{(1,x)}_-}$};
	\node[font=\LARGE] (C) at (0 : 3cm) {$\mathcolor{red}{\Psi^{(2,z)}_-}$};
	\node[font=\LARGE] (D) at (315 : 3cm) {$\bar{\Psi}^{(2,x)}_-$};
	\node[font=\LARGE] (E) at (270 : 3cm) {$\bar{\Psi}^{(1,z)}_-$};
	\node[font=\LARGE] (F) at (225 : 3cm) {$\bar{\Psi}^{(1,x)}_-$};
	\node[font=\LARGE] (G) at (180 : 3cm) {$\bar{\Psi}^{(2,z)}_-$};
	\node[font=\LARGE] (H) at (135 : 3cm) {$\mathcolor{blue}{\Psi^{(2,x)}_-}$};
	\draw [->, thick] (A) to [bend left = 20] (B);
	\draw [->, thick] (B) to [bend left = 20] (C);
	\draw [->, thick] (C) to [bend left = 20] (D);
	\draw [->, thick] (D) to [bend left = 20] (E);
	\draw [->, thick] (E) to [bend left = 20] (F);
	\draw [->, thick] (F) to [bend left = 20] (G);
	\draw [->, thick] (G) to [bend left = 20] (H);
	\draw [->, thick] (H) to [bend left = 20] (A);
\end{tikzpicture}

Here we use the following notation, where the maps to (unormalized) states in the free vector space are indicated:
\begin{align*}
	\Psi^{(1,z)}_- &= \frac{1}{\sqrt{2}} \Big[ |0\rangle_A \otimes |1\rangle_B + |-1\rangle_A \otimes |0\rangle_B  \Big] \\
	&\rightarrow [ (e_0, f_1) + (-e_1, f_0) ] \\
	\Psi^{(2,z)}_- &= \frac{1}{\sqrt{2}} \Big[ |0\rangle_A \otimes |1\rangle_B + |1\rangle_A \otimes |-0\rangle_B  \Big] \\
	&\rightarrow [ (e_0, f_1) + (e_1, -f_0) ]  \\
	\Psi^{(1,x)}_- &= \frac{1}{\sqrt{2}} \Big[ |1'\rangle_A \otimes |0'\rangle_B + |0'\rangle_A \otimes |-1'\rangle_B  \Big] \\
	&\rightarrow  [ (e_0 - e_1, f_0 + f_1) + (e_0 + e_1, -f_0 + f_1) ] \\
	\Psi^{(2,x)}_- &= \frac{1}{\sqrt{2}} \Big[ |1'\rangle_A \otimes |0'\rangle_B + |-0'\rangle_A \otimes |1'\rangle_B \Big] \\
	&\rightarrow  [ (e_0 - e_1, f_0 + f_1) + (-e_0 - e_1, f_0 - f_1) ]  \\
	\bar{\Psi}^{(1,z)}_- &= \frac{1}{\sqrt{2}} \Big[ |-0\rangle_A \otimes |-1\rangle_B + |1\rangle_A \otimes |-0\rangle_B  \Big] \\
	&\rightarrow  [ (-e_0, -f_1) + (e_1, -f_0) ]  \\
	\bar{\Psi}^{(2,z)}_- &= \frac{1}{\sqrt{2}} \Big[ |-0\rangle_A \otimes |-1\rangle_B + |-1\rangle_A \otimes |0\rangle_B  \Big] \\
	&\rightarrow [ (-e_0, -f_1) + (-e_1, f_0) ]   \\
	\bar{\Psi}^{(1,x)}_- &= \frac{1}{\sqrt{2}} \Big[ |-1'\rangle_A \otimes |-0'\rangle_B + |-0'\rangle_A \otimes |1'\rangle_B  \Big] \\
	&\rightarrow  [(-e_0 + e_1, -f_0 - f_1) + (-e_0 - e_1, f_0 - f_1) ]  \\
	\bar{\Psi}^{(2,x)}_- &= \frac{1}{\sqrt{2}} \Big[ |-1'\rangle_A \otimes |-0'\rangle_B + |0'\rangle_A \otimes |-1'\rangle_B \Big] \\
	&\rightarrow   [(-e_0 + e_1, -f_0 - f_1) + (e_0 + e_1, -f_0 + f_1) ]   .
\end{align*}

The cycle of basis transformations of a wavefunction, such as illustrated above, shows the full structure of an entangled state. Not only does the wavefunction encode various basis-dependent joint measurement outcomes, it encodes also the correlations and apparent randomness characteristic of these measurements. This is shown by explicitly rotating the basis and seeing that we cycle through not only basis representations, but alternative phase classes associated with those representations.  It is evident that the \emph{same} wavefunction can exhibit both class 1 and 2 contextual phases classes, just like the same wavefunction can be measured in different bases. This shows that the outcomes of measurements are not predetermined, or decided by local variables. 

We will now show why the joint measurement correlations do not depend on the contextual phase class. For arbitrary measurement settings denoted by detection ``angle'' $\alpha$ for subsystem A and $\beta$ for subsystem B we use $U = R_{\alpha}  \otimes R_{\beta} $, defined above. We may either rotate the basis of the wavefunction and evaluate:
\begin{equation}
	\langle \alpha \beta \rangle = \langle U \Psi | \sigma_z \otimes \sigma _z | U \Psi \rangle,
\end{equation}
or we can rotate the operator and evaluate
\begin{align}
	\langle \alpha \beta \rangle &= \langle \Psi | U^{\dagger} \sigma_z \otimes \sigma _z  U|  \Psi \rangle \\
	&= \langle \Psi |  \sigma_{\alpha} \otimes \sigma _{\beta}  |  \Psi \rangle \nonumber .
\end{align}
As shown earlier in this section, applying the rotation operator to the subsystems rotates their local measurement basis as well as contextual phase class. Whereas, the equivalence of Eqs 20 and 21 for evaluating joint measurement correlations indicates that the joint correlations in measurement outcomes are independent of contextual phase class. It is only the specific measurement outcomes that depend on contextual phase. 

This result can be shown clearly with a specific example. Consider 
\begin{equation*}
	\Psi_- = \frac{1}{\sqrt{2}} \Big( |0\rangle \otimes |1\rangle - |1\rangle \otimes |0\rangle  \Big),
\end{equation*}
and we will evaluate $\langle \alpha \beta \rangle_{\Psi_-} = \langle U \Psi_- | \sigma_z \otimes \sigma _z | U \Psi_- \rangle$. It will be shown why the evaluation is independent of whether we write $\Psi_-$ in the tensor product space (as above) or whether we consider  representations of $\Psi_-$ with explicit class 1 and 2 phases, that is, $\Psi^{(1,z)}_-$ and $\Psi^{(2,z)}_-$. All we need to do is to compute the matrix element. We can do this for states explicitly in the free vector space or, as we choose to show here, we will evaluate in the tensor product space but carefully point out what happens to the phase of the superposition for the states $\Psi^{(1,z)}_-$ and $\Psi^{(2,z)}_-$ in the free vector space. 

We will use the action of $\sigma_z$ on the basis vectors:
\begin{equation*}
	\sigma_z|0\rangle = |0\rangle, \qquad \sigma_z|1\rangle=-|1\rangle,
\end{equation*}
and the action of the rotation operator $R_{\theta}$ on the basis vectors:
\begin{eqnarray*}
	R_{\theta} |0\rangle = \cos \theta |0\rangle + \sin \theta |1\rangle \\
	R_{\theta} |1\rangle = -\sin \theta |0\rangle + \cos \theta |1\rangle.
\end{eqnarray*}

First evaluate
\begin{align}
	R_{\alpha} \otimes R_{\beta} |\Psi_- \rangle &= \frac{1}{\sqrt{2}} \Big[ R_{\alpha} |0\rangle \otimes  R_{\beta} |1\rangle - R_{\alpha} |1\rangle \otimes  R_{\beta} |0\rangle  \Big] \\
	&= \frac{1}{\sqrt{2}} \Big[ R_{\alpha} |0\rangle \otimes  R_{\beta} |1\rangle + R_{\alpha} (-1)|1\rangle \otimes  R_{\beta} |0\rangle  \Big]   \\ \nonumber
	&= \frac{1}{\sqrt{2}} \Big[ R_{\alpha} |0\rangle \otimes  R_{\beta} |1\rangle + R_{\alpha} |1\rangle \otimes  R_{\beta} (-1) |0\rangle  \Big] . \nonumber
\end{align}
Here we see that the result will not depend on whether we compute the matrix element for $\Psi_-$, $\Psi^{(1,z)}_-$, or $\Psi^{(2,z)}_-$. That is not surprising because we must get the same result by the universal mapping property. That shows that the correlations between joint measurement outcomes do not depend on the contextual phase class, which is what we set out to examine. We continue to compute the correlation simply for completeness. 

\begin{eqnarray*}
	R_{\alpha} \otimes R_{\beta} |\Psi_- \rangle &= \frac{1}{\sqrt{2}} \Big[ (-\cos \alpha \sin \beta + \sin \alpha \cos \beta) |00\rangle \\
	&+ (\sin \alpha \cos \beta - \cos \alpha \sin \beta) |11\rangle \\
	&+ (\cos \alpha \cos \beta + \sin \alpha \sin \beta) |01\rangle \\
	&+ (-\sin \alpha \sin \beta - \cos \alpha \cos \beta) |10\rangle \Big] \\
	&= \frac{1}{\sqrt{2}} \Big[ \sin (\alpha - \beta)(|00\rangle + |11\rangle) + \cos (\alpha - \beta)(|01\rangle + |10\rangle) \Big].
\end{eqnarray*}
The final step used the trigonometric product to sum identities. Now evaluate the matrix element
\begin{eqnarray}
	\langle \alpha \beta \rangle_{\Psi_-} &= \langle R_{\alpha} \otimes R_{\beta} \Psi_- | \sigma_z \otimes \sigma _z | R_{\alpha} \otimes R_{\beta} \Psi_- \rangle \\
	&= \sin^2(\alpha  - \beta) - \cos^2(\alpha  - \beta) \\ \nonumber
	&= -\cos 2(\alpha  - \beta) . \nonumber
\end{eqnarray}

Therefore, the correlations among joint measurement outcomes do not depend on the contextual phase class for similar reasons that the correlations do not depend on the basis we choose to represent the state. We showed the explicitly example here of $\langle \alpha \beta \rangle_{\Psi_-}$. Ultimately what is important is that the correlations come from evaluation of the inner product. A brief discussion is given in Appendix C.

\section{Discussion}
In this paper we aimed to clarify the following questions in particular: (a) What are measurements on separated subsystems? (b) If the other particle in the pair is not influenced in any way by measurements on the first particle, how are the measurement outcomes correlated?  We proposed that these questions can be answered by defining more deeply what are separated subsystems and how they give a projection of the entangled state. Fig. 4 summarizes the model concluded from our work. Definite outcomes of measurements are explained by understanding how measurements on the subsystems are instrumental in producing the definite outcomes from the superpositions being measured. This problem is intimately related to measurements of separated subsystems in general, and thereby gives insight into the origin of nonlocal correlations. 

\begin{figure}[h]
	\centering
	\includegraphics[width=0.8\textwidth]{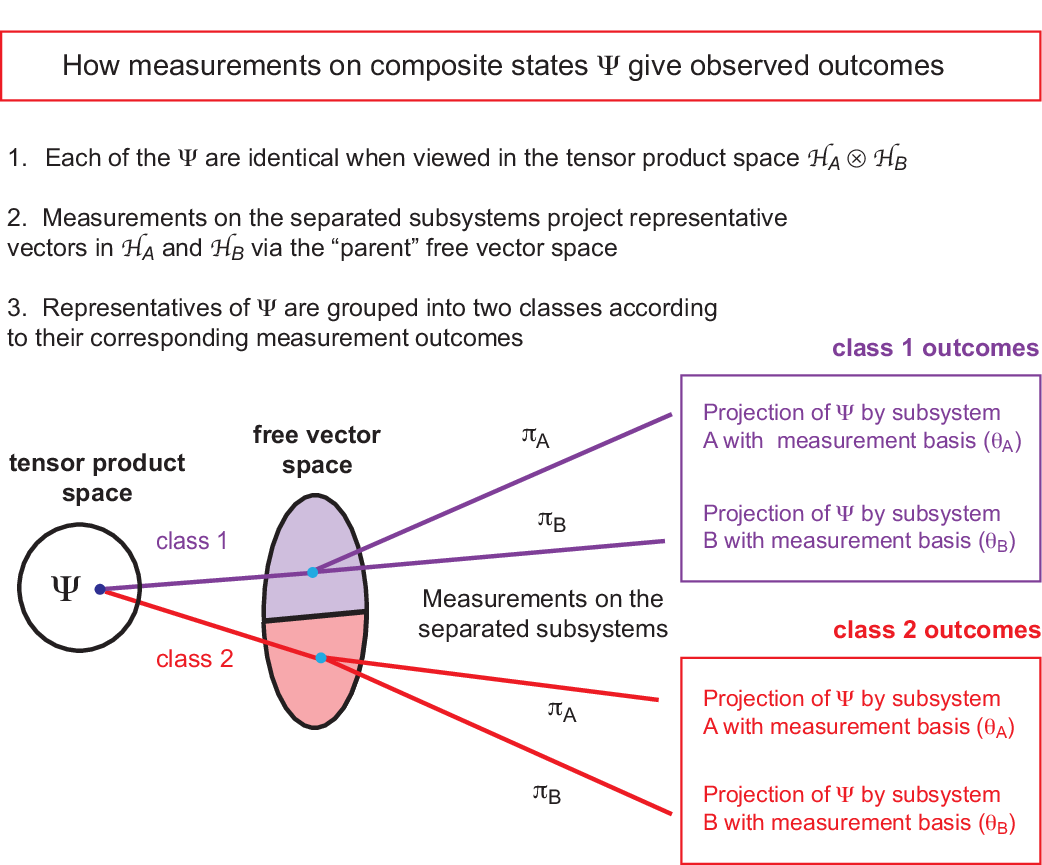}
	\caption{Summary of how contextual phase theory translates a composite state $\Psi$ to definite measurement outcomes on the separated subsystems.  }\label{fig4}
\end{figure}

Our results can be given in the context of one of the programs initiated by the EPR paper\cite{EPR}. An important outcome of the EPR paper has been to highlight the perplexing implications of the collapse postulate (in the form stated above) for measurements on the separated subsystems of entangled states. Bohm's EPR state provides a clear platform for interpreting the issue:
\begin{equation}
	\Psi_- = \frac{1}{\sqrt{2}} \Big( |0\rangle_A |1\rangle_B - |1\rangle_A |0\rangle_B  \Big),
\end{equation}
where $|0\rangle_A$ and $|1\rangle_A$ are basis vectors for subsystem A in $\mathcal{H}_A$, etc.  The key step in the EPR paper is to consider measurements on the subsystems after they have been separated from one another. Now, we know that $\langle \Psi_- | \sigma_z \otimes \sigma_z | \Psi_- \rangle = -1$ (where $\sigma_z$ is a Pauli operator), meaning that the outcomes of measurements in the $z-$basis on the separated subsystems must be anticorrelated. For example, if subsystem A has eigenvalue $-1$ (0 in our notation), then subsystem B has eigenvalue 1. Yet, a naive inspection of Eq 26, together with knowledge of the collapse rule, suggests no reason why a measurement on subsystem A---randomly either 0 or 1 according to the random collapse of the superposition---should be correlated to a measurement on subsystem B. Something is missing from the theory underpinning the measurement postulate. Otherwise we are forced to speculate how the measurement outcome on subsystem A is, somehow, instantaneously communicated to subsystem B so as to correlate the measurement outcomes. The issue has led to the development of many varied interpretations for quantum-mechanical measurements that bypass or rationalize the paradox\cite{Laloe, Dickson, QBism1, QBism2, QBism3, Zwirn2023, CSM, Fine1982, Mermin1985, Vaidman2014}. The theory of quantum mechanics is beyond doubt, but it is widely recognized that this measurement problem needs a compelling resolution. The present paper makes the case that contextual phase theory fills this gap.

One of the most interesting and much debated questions is how the measurement outcomes for the two subsystems are correlated. As stated in Ref. \cite{Einsteinstruggles} ``Einstein had no difficulty accepting that affairs in different places could be correlated. What he could not accept was that an intervention at one place could \emph{influence} immediately affairs at another''. Contextual phase theory shows how the superpositions of tensors in the composite state $\Psi$ lead to phases in the projections of $\Psi$ to the subsystems that encode correlations. These correlations are embedded in the state vector (although hidden in the tensor product representation) and are simply revealed by measurements on the separated subsystems---there is no `spooky action at a distance'. However, the deep aspect of the issue comes from the fact that the measurement outcomes are undecided until the measurement basis is chosen. The additional factor we propose here is that the measurement basis includes contextual phase class. So, even though contextual phases encode the correlations, their implications depend on the measurement basis. 

The measurement postulates are generally thought to imply the `collapse' rule, which intuitively resolves the problem of obtaining a physical measurement outcome from a superposition state.  What do contextual phases tell us about collapse? Have we simply moved the randomness somewhere else?  No, instead the randomness is replaced by a statistical basis for the measurement outcomes because a single wavefunction $\Psi$ in the tensor product basis is associated, by the quotient, to multiple classes of measurable \emph{separated} states in the free vector space. Grouping the representatives for $\Psi$ into different contextual phase classes shows how measurement outcomes expose, statistically, any one of those classes. This analysis suggests how a \emph{single} $\Psi$ gives the kind of ensemble view of a wavefunction imagined by Einstein\cite{FineBook, Werner2014}.

The random element of collapse is transferred to a special kind of \emph{statistical} distribution of contextual phase class that directs measurement outcomes for the entangled state of the separated subsystems. Collapse \emph{appears to be} random because the contextual phase class is statistically distributed among the representatives of the entangled state in the covering free vector space. That is, all the wavefunctions are identical, but contextual phase class is statistically presented. This new viewpoint was enabled by proposing a theory for measurements on the separated subsystems A and B of entangled states; essentially resolving what are the local vectors in $\mathcal{H}_A$ and $\mathcal{H}_B$ and how they are associated to the entangled state.

A contextual phase class applies to the entire composite system and therefore collapse of a state vector for one subsystem in a composite state is correlated with the collapse of the state vector of the other subsystem, without the need for any kind of inexplicable long-range interaction after the subsystems separate. The correlations derive from the fact that phases incorporated in the subsystem states are locked together by the phase in the superposition of a composite state.

\section{Conclusions}

We have focused on the measurements of observables of separated subsystems from bipartite entangled states, with Einstein's definition in mind: ``Separation is the claim that whether a physical property holds for one of the particles does not depend on measurements (or other interactions) made on the other particle when the pair is widely separated in space.'' The work clarifies tow main questions: (a) What are measurements on separated subsystems? (b) If the other particle in the pair is not influenced in any way by measurements on the first particle, how are the measurement outcomes correlated? The developments reported here build on prior work\cite{collapse} that proposes how to map entangled states in $\mathcal{H}_A \otimes \mathcal{H}_B$ to states that represent the appropriate local superpositions in the separated, non-interacting subsystems. Those states lie in $\mathcal{H}_A$ and $\mathcal{H}_B$.

The main result of the paper is that contextual phase theory explains why definite measurement outcomes happen within the framework of quantum mechanics. This concept gives a plausible mechanism for explaining collapse in the quantum measurement postulates. Moreover, it guarantees that when a measurement is performed, only one result is obtained and repeated measurements give the same result.  The theory makes these predictions without assuming any kind of \emph{ad hoc} collapse. That is, we propose an explanation for the Copenhagen interpretation of measurement. 

If the present manuscript is to be summarized by three key concepts, they are: (i) Projections of the entangled state enabled by separated subsystems come from the free vector space $U \times V$, not the quotient space $U \otimes V$; (ii) The different ways the phase of the superposition appears in $U \times V$ controls the measurement outcome (collapse is an interference effect); (iii) A single wave function therefore encodes an ensemble of measurement outcomes.




\noindent


\vspace{6pt} 




\bibliographystyle{tfq}
\bibliography{Scholes_bib_Nov2026}

\section{Appendices}

\noindent\textbf{Appendix A. Free vector spaces}\medskip

\begin{definition}{(Cartesian product of sets)}
	Let $S_1, \dots, S_n$ be any sets. The Cartesian product of sets, denoted $S_1 \times \dots \times S_n$, is the set of all ordered $n-$tuples $(a_1, \dots, a_n)$, where $a_i \in S_i$ for $i = 1, \dots n$.
\end{definition}

\begin{definition}{(Free vector space)}
	A free vector space $\mathcal{F}(\mathcal{S})$ on a set $\mathcal{S}$ over a field $\mathbb{K}$ is a vector space comprising all finite formal linear combinations of elements in $\mathcal{S}$, where the elements of $\mathcal{S} = \{ e_0, e_1, \dots \}$ form a basis.
\end{definition}

In the present work we define $\mathcal{F}(U \times V)$ as the free vector space over the field $\mathbb{C}$ on the set $U \times V$ (i.e. Cartesian product). It has elements that are finite $\mathbb{C}$-linear combinations of basis symbols and satisfies the universal property for maps from $U \times V$ into $\mathbb{C}$-vector spaces. 

The notion of formal linear combinations means that we have linear combinations of the form $\alpha_0 e_0 + \alpha_1 e_1 + \dots$ ($\alpha_i \in \mathbb{K}$), but no relations among the symbols (i.e. $e_0, e_1, \dots$) are assumed. That is, the symbols are simply elements of a set (e.g. they could be red, blue, green, etc.). That is fundamentally different than if they were basis vectors in Cartesian space, for example. However, by defining a free vector space, we endow these elements of a set with the properties of a vector space\cite{Roman}. For instance, we have 
\begin{equation*}
	r e_i + s e_i = (r + s)e_i
\end{equation*}
and
\begin{equation*}
	r(s e_i) = (rs) e_i .
\end{equation*}

Thus, for example, $3e_0 -2e_0 + e_1 = e_0 + e_1$. We can easily map the elements of $\mathcal{F}(\mathcal{S})$ to $\mathcal{H}$, then in the vector space $\mathcal{H}$ the formal sums from $\mathcal{F}(\mathcal{S})$ can be collapsed in the same way: $3e_0 -2e_0 + e_1 \rightarrow 3|0\rangle - 2|0\rangle + |1\rangle$, that simplifies to $ |0\rangle + |1\rangle$. 

To clarify the nature of this map $\mathcal{F} \rightarrow \mathcal{H}$, we define a free vector space $\mathcal{F}(U)$ such that its elements map bijectively to the Hilbert space vectors of a quantum subsystem $\psi = \alpha|0\rangle + \beta|1\rangle$. That is, we take our basis to be $\mathbb{Z}_2 = \{0, 1\}$, written $\{e_0, e_1\}$, and define the map $e_0 \rightarrow |0\rangle$ and $e_1 \rightarrow |1\rangle$. Define $\mathcal{F}(U)$ over $\mathbb{C}$ as all finite $\mathbb{C}$-linear combinations of the formal basis functions $e_0$ and $e_1$ so that $\mathcal{F}(U) \cong \mathbb{C}^2$ with elements $\alpha e_0 + \beta e_1$. Choose the identification map $T: \mathcal{F}(U) \rightarrow \mathcal{H}$ such that $T(e_0) = |0\rangle$ and $T(e_1) = |1\rangle$, where $|0\rangle$ and $|1\rangle$ form the basis of $\mathcal{H}$. Then extend linearly, so that $T(\alpha e_0 + \beta e_1) =  \alpha|0\rangle + \beta|1\rangle$. 

In the present work we encounter wavefunctions that we write in a way to specify the state in the overlying free vector space. For instance 
\begin{equation*}
	\Psi_+ = \frac{1}{\sqrt{2}} \Big[ |0'\rangle_A \otimes |0'\rangle_B + |1'\rangle_A \otimes |-1'\rangle_B  \Big] ,
\end{equation*}
that corresponds to
\begin{equation*}
	\Psi_+ = \frac{1}{\sqrt{2}} \Big[ (e_0 + e_1, e_0 + e_1) +  (e_0 - e_1, -e_0 + e_1) \Big] .
\end{equation*}
Each of these two terms now is a generator for the free vector space Cartesian product. Unlike similar-looking points in the Cartesian plane, that can be added together component-wise, we cannot add these points. However, after we take the canonical projection map to one subsystem (say A), we obtain
\begin{equation}
	\pi_A(\Psi_+) = u_A = \frac{1}{2}  ( e_0 + e_1 + e_0 - e_1  ) .
\end{equation}
Now this vector is a sum of generators $e_0$ and $e_1$ that can be simplified, giving $u_A = e_0$. Note, however, that this simplification of the projected vector only happens when a measurement is performed---otherwise entanglement would be destroyed solely by separating the subsystems.

\vspace{24pt} 
\noindent\textbf{Appendix B Contextual phase is connected to measurement basis. }\medskip

Consider a pair of two-level systems A and B, where their two state vectors are labeled $a_1, a_2 \in \mathcal{H}_A$ and $b_1, b_2 \in \mathcal{H}_B$ respectively. For the first example we consider the following separable states written in pre-image form so they relate to the corresponding elements in the free vector space:
\begin{eqnarray*}
	\Theta_{++} = \frac{1}{2} (a_1 + a_2) \otimes (b_1 + b_2) \\
	\Theta_{--} = \frac{1}{2} (a_1 - a_2) \otimes (b_1 - b_2) \\
	\tilde{\Theta}_{--} = \frac{1}{2} (a_1 - a_2) \otimes (-b_1 + b_2) \\
	\tilde{\Theta}'_{--} = \frac{1}{2} (-a_1 + a_2) \otimes (b_1 - b_2) .
\end{eqnarray*}
Notice that $\Theta_{--} = \tilde{\Theta}_{--} = \tilde{\Theta}'_{--} $, where $\tilde{\Theta}_{--}$ and $\tilde{\Theta}'_{--}$ each include an embedded negative overall phase and $\tilde{\Theta}'_{--}$ is the antipodal vector to $\tilde{\Theta}_{--}$ in the space $\mathbb{S}^3_A \times \mathbb{S}^3_B$. 

The entangled states $\Psi_+$ can be constructed in various ways as an explicit sum of separable states, including:
\begin{align*}
	\Psi_+ &= \frac{1}{\sqrt{2}} (\Theta_{++} - \Theta_{--}) \\
	&= \frac{1}{\sqrt{2}} (\Theta_{++} + \tilde{\Theta}_{--}) \quad \text{class 1} \\
	&= \frac{1}{\sqrt{2}} (\Theta_{++} + \tilde{\Theta}'_{--}) \quad \text{class 2},
\end{align*}
which are all equivalent in the tensor product space $\mathcal{H}_A \otimes \mathcal{H}_B$ where they simplify by interference to $\frac{1}{\sqrt{2}} (a_1b_2 + a_2b_1)$. 

Consider the entangled state written as a pre-image to make clear the class 1 contextual phases: $\Psi_+^{(1,x)} = \frac{1}{\sqrt{2}} (\Theta_{++} + \tilde{\Theta}_{--})$. To be consistent, the wavefunctions in the $x$-basis form, that yield $z$-basis measurement outcomes, are indicated with a $x$ superscript. Upon projection to the separated subsystems we find
\begin{align*}
	\pi_A(\Psi_+^{(1,x)}) &= \frac{1}{2}(a_1 + a_2) + \frac{1}{2}(a_1 - a_2) \\
	&= a_1 \\
	\pi_B(\Psi_+^{(1,x)}) &= \frac{1}{2}(b_1 + b_2) + \frac{1}{2}(-b_1 + b_2) \\
	&= b_2.
\end{align*}

On the other hand, $\Psi_+^{(2,x)} = \frac{1}{\sqrt{2}} (\Theta_{++} + \tilde{\Theta}'_{--})$, where the vector carrying the phase of the superposition is taken to be the antipodal vector, gives class 2 measurement outcomes for the separated subsystems:
\begin{align*}
	\pi_A(\Psi_+^{(2,x)}) &= \frac{1}{2}(a_1 + a_2) + \frac{1}{2}(-a_1 + a_2) \\
	&= a_2 \\
	\pi_B(\Psi_+^{(2,x)}) &= \frac{1}{2}(b_1 + b_2) + \frac{1}{2}(b_1 - b_2) \\
	&= b_1.
\end{align*}

The measurement outcomes in the $z$-basis are anticorrelated and come in random sequence because whether $\Psi_+$ is class 1 or 2 is random. 

We can also obtain $x$-basis measurement outcomes on the separated subsystems for this state. Those outcomes are easily ascertained by inspecting the simplified vector $\Psi_+ = \frac{1}{\sqrt{2}} (a_1b_2 + a_2b_1)$. More interestingly, we can predict correlated measurement outcomes for a $z$-basis composite state (i.e. $x$-basis measurement outcomes)  from in- and out-of-phase linear combinations of projections into the Hilbert spaces of the separated subsystem $U$ for $x$-basis states and their contextual phases. This allows us to see how different measurement bases detect different linear combinations of contextual phase classes:
\begin{eqnarray}
	\pi_U(\Psi_+^{(1,z)}) = \frac{1}{\sqrt{2}}\pi_U(\Psi_+^{(1,x)} + \Psi_+^{(2,x)}) \\
	\pi_U(\Psi_+^{(2,z)}) = \frac{1}{\sqrt{2}}\pi_U(\Psi_+^{(1,x)} - \Psi_+^{(2,x)}),
\end{eqnarray}
Such that the class 1 outcomes are, for instance
\begin{align*}
	\pi_A(\Psi_+^{(1,z)}) &= \frac{1}{\sqrt{2}} (\pi_A(\Psi_+^{(1,x)}) + \pi_A(\Psi_+^{(2,x)}) ) \\
	&= \frac{1}{\sqrt{2}}(a_1 + a_2) \\
	\pi_B(\Psi_+^{(1,z)}) &= \frac{1}{\sqrt{2}} (\pi_B(\Psi_+^{(1,x)}) + \pi_B(\Psi_+^{(2,x)}) ) \\
	&= \frac{1}{\sqrt{2}}(b_1 + b_2) .
\end{align*}
The corresponding projections on the antisymmetric linear combinations yield the class 2 outcomes for the $x$-basis measurements $\frac{1}{\sqrt{2}}(a_1 - a_2)$ and $\frac{1}{\sqrt{2}}(b_1 - b_2)$. This computation shows that $x$-basis measurement outcomes are decided by linear combinations of contextual phase classes in the wavefunction that gives $z$-basis measurement outcomes---a unitary transformation of the phase, as expected.

We can repeat the procedure for other entangled states, for example by computing projections for $\Phi_-^{(1,x)} = \frac{1}{\sqrt{2}} (\Theta_{-+} + \Theta_{+-})$ and $\Phi_-^{(2,x)} = \frac{1}{\sqrt{2}} (\Theta_{-+} + \tilde{\Theta}_{+-})$, where
\begin{eqnarray*}
	\Theta_{-+} = \frac{1}{2} (a_1 - a_2) \otimes (b_1 + b_2) \\
	\Theta_{+-} = \frac{1}{2} (a_1 + a_2) \otimes (b_1 - b_2) \\
	\tilde{\Theta}_{+-} = \frac{1}{2} (-a_1 - a_2) \otimes (-b_1 + b_2)  .
\end{eqnarray*}
Here $\tilde{\Theta}_{+-}$ is the antipodal vector to $\Theta_{+-}$ in the space $\mathbb{S}^3_A \times \mathbb{S}^3_B$. 

In this exercise, $\pi_A(\Phi_-^{(2,x)}) = -a_2$. The overall phase does not matter here, but it needs to be retained for predicting the $x$-basis measurements using the procedure described above. Taking the information directly from Table 2, Eqs. 28 and 29 predict measurement outcomes in the $x$-basis from those in the $z$-basis, as summarized in Table 3.

\begin{table}[!h]
	\caption{Measurement outcomes on separated subsystems where the $x$-basis results are computed using Eqs. 28 and 29. }
	\label{table_3}
	\begin{tabular}{llll}
		\hline
		State & class & $z$-basis outcomes (A, B) & $x$-basis outcomes (A, B)  \\
		\hline
		$\Psi_+$ & 1 & $0 \qquad 1$ & $0' \qquad 0'$   \\
		& 2 & $1 \qquad  0$ & $1' \qquad  -1'$ \\
		$\Phi_+$ & 1 & $0 \qquad  0$ & $0' \qquad  0'$   \\
		& 2 & $1 \qquad  1$ & $1' \qquad  1'$ \\
		$\Phi_-$ & 1 & $0 \qquad  0$ & $1' \qquad  0' $  \\
		& 2 & $-1 \qquad  1$ & $0' \qquad  1'$ \\
		$\Psi_-$ & 1 & $0 \qquad  1$ & $1' \qquad  0'$   \\
		& 2 & $-1 \qquad  0$ & $0' \qquad  -1'$ \\
		\hline
	\end{tabular}
	\vspace*{-4pt}
\end{table}

\vspace{24pt} 
\noindent\textbf{Appendix C Quantum correlations that violate Bell's inequality}\medskip

More formally, why the quantum correlations can exceed the best possible classical correlations can be understood through the lens of Grothendieck's inequality\cite{Grothendieck, Pisier2011, DFS}. The connection between Grothendieck's inequality and Bell's inequality was first noticed by Tsirelson\cite{Tsirelson1, FR1994}. Grothendieck's inequality comes in several different forms\cite{DFS, Diestel1984, AlbiacKalton, DefantFloret, DJT}. A relevant (operator-perspective) version is stated as follows\cite{DJT}:

\begin{theorem}{(Grothendieck's Inequality)}
	There is a universal constant $K_G$ for which, given any Hilbert space $\mathcal{H}$ and any $n \in \mathbb{N}$, any $n \times n$ scalar matrix $(a_{ij})$ and any vectors $x_1, \dots, x_n$ and $y_1, \dots, y_n$ in the Banach space $B_{\mathcal{H}}$, we have
	\begin{equation}
		\Big| \sum_{i,j} a_{ij} \langle x_i | y_j \rangle  \Big| \le K_G \max \Big\{ \Big| \sum_{i,j} a_{ij} s_i t_j  \Big| : |s_i| \le 1, |t_j| \le 1 \Big\} .
	\end{equation}
\end{theorem}

Here, the matrix $A = (a_{ij})$ comprises elements $a_{ij}$ that act as weights or coefficients for the specific combination of measurement settings indexed by $i$ and $j$. For example, these can be chosen as the coefficients from the CHSH inequality.  The sequences $s_i$ and $t_j$ represent trials of definite measurement outcomes for each subsystem, restricted to classical scalar values of $\pm 1$. By optimizing these sequences, the right-hand side of Eq. 30 gives the maximum classical expectation value (the classical bound) for the sum of the product of coefficients from the Bell functional $a_{ij}$ with $s_i t_j$.  

Understanding the left-hand side of Eq. 30 takes more work. To capture the quantum correlations on the left-hand side, we replace the scalar outcomes with vectors in a Hilbert space. The classical measurement setting $i$ for subsystem A (an element of the Banach space $\ell_1$) is mapped to a measurement vector in the local space $\mathcal{H}_A$. To evaluate the joint correlation, we must bring this vector and the corresponding vector for subsystem B into a common Hilbert space. This is accomplished using the canonical isomorphism (Chapter 2 of Ref. \cite{DefantFloret}, Chapter 1 of \cite{DFS}):
\begin{equation}
	\mathcal{H}_A \otimes \mathcal{H}_B \cong \mathcal{L}(\mathcal{H}^*_A, \mathcal{H}_B),
\end{equation}
where $\mathcal{L}( \cdot )$ means the space of linear maps and $*$ denotes the dual space. This tells us that a state $\Psi \in \mathcal{H}_A \otimes \mathcal{H}_B$ is equivalent to a linear operator
\begin{equation}
	T_{\Psi} : \mathcal{H}^*_A \rightarrow \mathcal{H}_B.
\end{equation}

By mapping the measurement vector for separated subsystem A (viewed as an element of the dual space $\mathcal{H}_A^*$) by $T_{\Psi}$, it is mapped to the Hilbert space of the second separated subsystem $\mathcal{H}_B$. The joint correlation is then simply the inner product $\langle x_i | y_j \rangle$ evaluated in this common space. Grothendieck's theorem shows that the quantum optimization can give correlations that exceed the optimal classical strategy by at most the universal constant $K_G$. Contextual phase theory does not change these correlations, it simply explains the set of definite measurement outcomes obtained by the experiment. 

\appendix

\end{document}